\documentclass[]{interact}

\usepackage{epstopdf}
\usepackage{subfigure}

\usepackage[numbers,sort&compress,merge]{natbib}
\bibpunct[, ]{(}{)}{,}{n}{,}{,}

\newcommand\Tstrut{\rule{0pt}{2.6ex}}       

\theoremstyle{plain}

\theoremstyle{definition}

\theoremstyle{remark}

\usepackage{amsmath,amssymb,graphicx}

\usepackage{dcolumn}
\usepackage{bm}
\usepackage{multirow}

\usepackage{textcomp}
\usepackage[caption=false]{subfig}
\usepackage{ragged2e}
\usepackage{braket}
\renewcommand\[{\begin{equation}}
\renewcommand\]{\end{equation}}
\newcolumntype{L}[1]{>{\RaggedRight\hspace{0pt}}p{#1}}
\newcolumntype{R}[1]{>{\RaggedLeft\hspace{0pt}}p{#1}}
\begin{document}

\articletype{}

\title{Joint Spectral Characterization of Photon-Pair Sources}

\author{
\name{Kevin Zielnicki\textsuperscript{a},
Karina Garay-Palmett\textsuperscript{b},
Daniel Cruz-Delgado\textsuperscript{c},
Hector Cruz-Ramirez\textsuperscript{c},
Michael  F. O'Boyle\textsuperscript{a},
Bin Fang\textsuperscript{a},
Virginia O. Lorenz\textsuperscript{a},
Alfred B. U'Ren\textsuperscript{c}, and
Paul G. Kwiat\textsuperscript{a}\thanks{CONTACT Paul Kwiat. Email: kwiat@illinois.edu}}
\affil{\textsuperscript{a}Department of Physics, University of Illinois at Urbana-Champaign, 1110 W. Green Street, Urbana, IL 61801, USA;
\textsuperscript{b}Departamento de \'Optica, Centro de Investigaci\'on Cient\'ifica y de Educaci\'on Superior de Ensenada, Apartado Postal 360, 22860 Ensenada, Mexico;
\textsuperscript{c}Instituto de Ciencias Nucleares, Universidad Nacional
Aut\'onoma de M\'exico, Apartado Postal 70-543, 04510 Ciudad de Mexico, Mexico}
}

\maketitle

\begin{abstract}
The ability to determine the  joint spectral properties of photon pairs produced by the processes of spontaneous parametric downconversion (SPDC) and spontaneous four wave mixing (SFWM) is crucial for guaranteeing the usability of heralded single photons and polarization-entangled pairs for multi-photon protocols.  In this paper, we compare six different techniques that yield either a characterization of the joint spectral intensity or of the closely-related purity of heralded single photons. These six techniques include: i) scanning monochromator measurements, ii) a variant of Fourier transform spectroscopy designed to extract the desired information exploiting a resource-optimized technique, iii) dispersive fibre spectroscopy, iv) stimulated-emission-based measurement, v) measurement of the second-order correlation function $g^{(2)}$ for one of the two photons, and vi) two-source Hong-Ou-Mandel interferometry. We discuss the relative performance of these techniques for the specific cases of a SPDC source designed to be factorable 
and SFWM sources of varying purity, and compare the techniques' relative advantages and disadvantages.
\end{abstract}


\section{Introduction to Joint Spectral Characterization \label{sec:Introduction-Sources}}

Pairs of polarization-entangled photons are a critical resource for optical quantum information processing. The nonlinear optical processes of spontaneous parametric downconversion (SPDC) and spontaneous four wave mixing (SFWM) are commonly used to produce photon pairs, usually with additional (and often undesirable) correlations in frequency and transverse momentum \cite{grice_eliminating_2001,uren_photon_2003}. Correlations between the signal and idler photons cause the detection of one of the  photons in a given  pair to herald its partner into a mixed state, which inhibits interference between independent sources. For example, a key building block of a scalable quantum communication network, the quantum repeater, requires interfering photons from a series of independent sources \cite{Briegel1998,Duan2001,Zhao2003}. Joint spectral measurement is an important tool for characterizing and optimizing the behavior of such sources \cite{Harder2013}.

In this work we describe, implement and compare six techniques that yield either a characterization of the joint spectral intensity or of the closely-related purity of heralded single photons, in order to provide an in-depth overview that highlights the purposes for which each technique is suited, demonstrates the procedures required to implement each technique, and presents a quantitative comparison of the information the techniques can provide. This section provides a brief introduction to the principles of joint spectral measurement, while in the next two sections we discuss the details of specific measurement techniques. Section \ref{chap:measuring-joint-spectrum} describes four independent measurements of the joint spectrum, using scanning monochromators, two-dimensional Fourier transform spectroscopy, dispersive fibre spectroscopy and stimulated-emission-based measurement. Section \ref{chap:measuring-purity} discusses measurements based on correlation functions and two-source Hong-Ou-Mandel (HOM) interference; 
while these last techniques do not provide a direct means to visualize the joint spectrum, they do directly relate to the heralded single-photon purity, which is often the metric of interest. The field of joint spectral characterization is quickly evolving, and thus the methods covered here do not include more recently developed techniques (see, for example, \cite{Mittal2017}), and the quoted rates and acquisition times represent those achievable with commonly-used silicon avalanche photodiode detectors rather than high-efficiency superconducting detectors (see, for example, \cite{Gerrits2015}). In the conclusions we summarize the experimental results and quantify figures of merit based on our implementations.

\subsection{A Simple Model\label{sec:Simple-Model}}

To illustrate the use of joint spectral measurement, we first consider a simplified
version of the joint spectrum of photon pairs. This
model motivates the more complete theory and provides an intuition for how to think about the joint spectrum.

Suppose that the only constraint on the SPDC process, in which a pump photon of frequency $\omega_{p}$ is converted into a pair of `signal' and `idler' photons (at $\omega_s$ and $\omega_i$, respectively), is conservation
of energy, i.e., $\omega_{p} = \omega_{s}+\omega_{i}$. (For SFWM, instead of just one pump photon, two are annihilated to create the signal and idler photons; 
for degenerate pump photons, the following discussion can be applied with the energy conservation condition $2\omega_p = \omega_{s} + \omega_{i}$.).
If the pump spectral amplitude is described by a Gaussian function $A(\omega_p)$ centred at  frequency $2\omega_{0}$, with bandwidth $\sigma$, we can write

\begin{equation}
A(\omega_p)=M\exp\left[-\frac{(\omega_p-2\omega_{0})^{2}}{2\sigma^{2}}\right],\label{eq:pump-envelope}
\end{equation}

\noindent where $M$ is a normalization constant to preserve unit area of $| A(\omega_p)| ^2$. To simplify
the analysis, we can redefine the pump spectral amplitude in terms of $\nu_{p} \equiv \omega_{p}-2\omega_{0}$, such that it has a mean of zero. This applies to the signal and idler frequencies as well, where for degenerate signal and idler photons we define $\nu_{s}\equiv \omega_{s}-\omega_{0}$ and $\nu_{i}\equiv \omega_{i}-\omega_{0}$.
Substituting these definitions into Eq. \ref{eq:pump-envelope}, we can write the joint amplitude of the photon pairs as

\begin{equation}
\begin{split}
f(\nu_{s},\nu_{i}) &=A(\omega_p=\omega_s+\omega_i)\\
&=M\exp\left[-\frac{\nu_{s}^{2}+\nu_{i}^{2}}{2\sigma^{2}}-\frac{2\nu_{s}\nu_{i}}{2\sigma^{2}}\right].\label{eq:simple-joint-spectrum}
\end{split}
\end{equation}

\begin{figure}
\centering
(a)\includegraphics[width=0.90in]{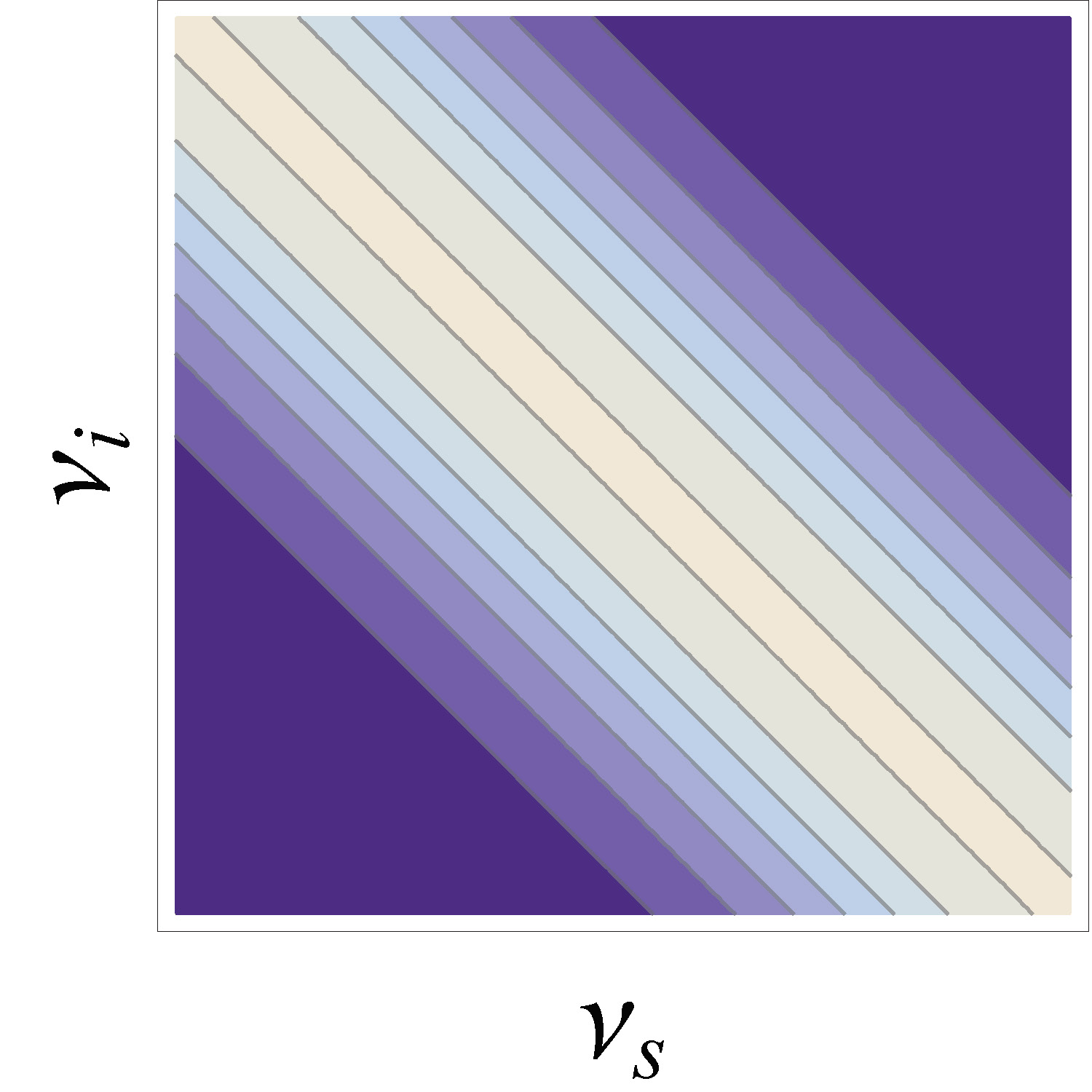}(b)\includegraphics[width=0.95in]{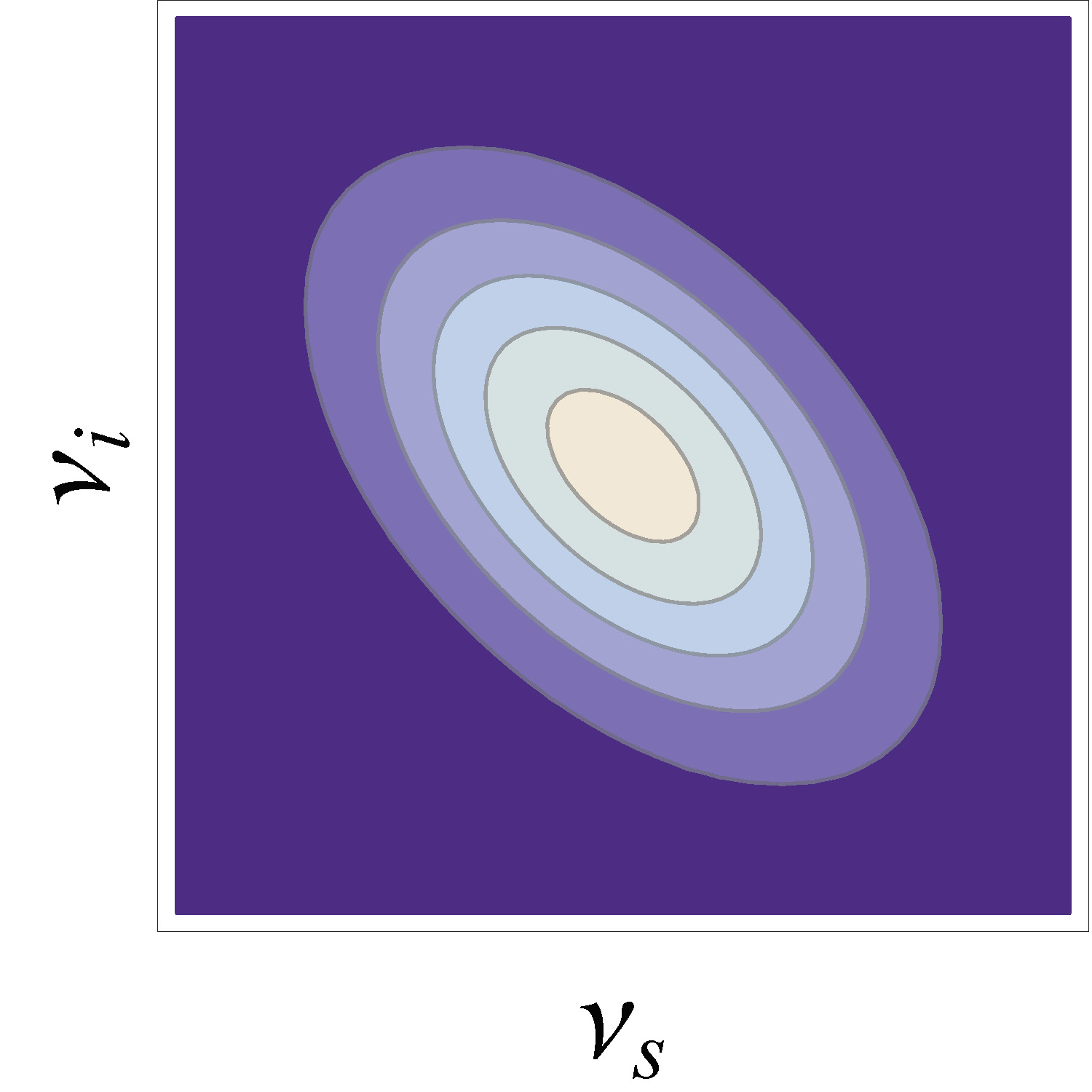}(c)\includegraphics[width=1in]{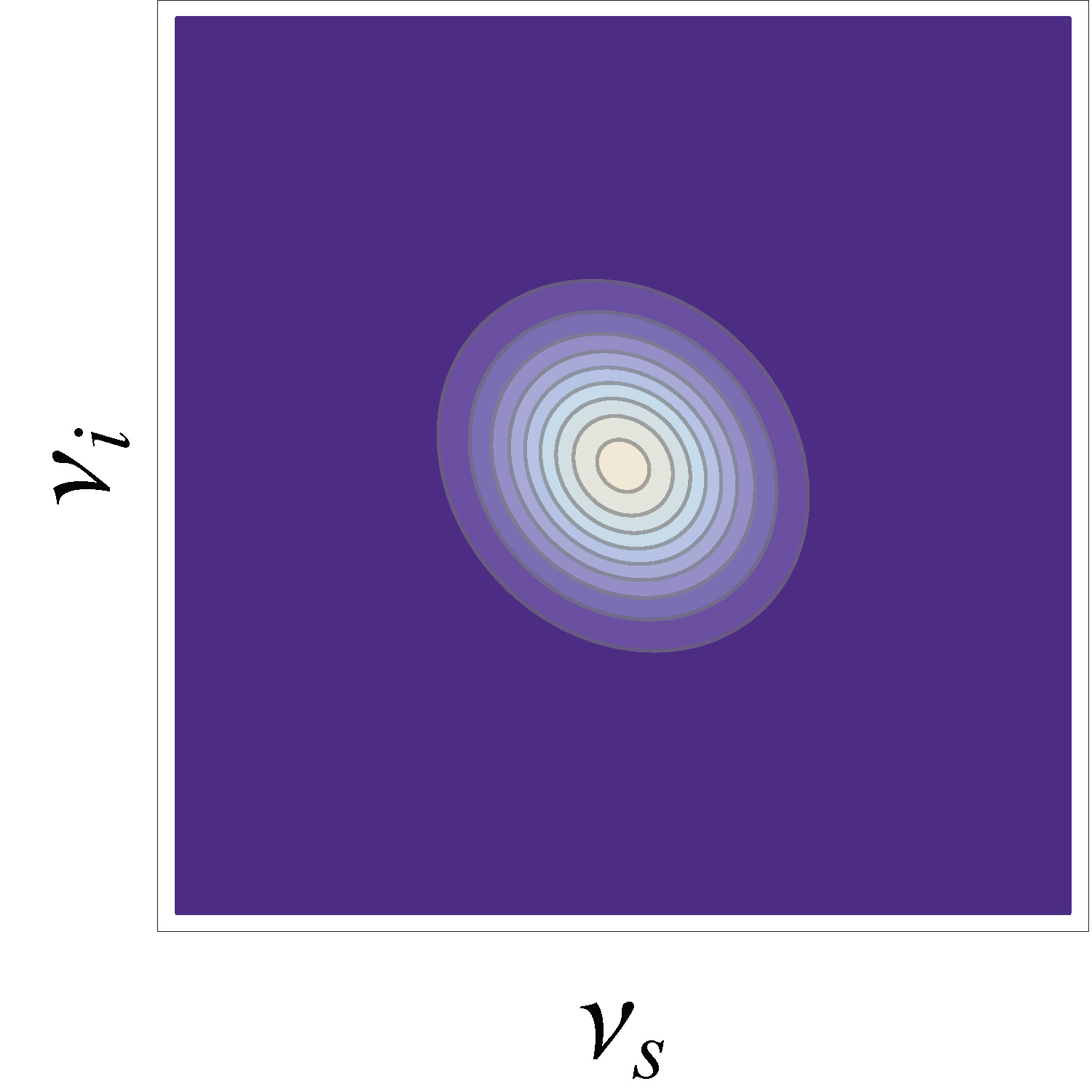}

\caption[Simple filtered JSI model]{Modeled JSI with (a) no filtering (Eq. \ref{eq:simple-joint-spectrum}),
(b) filter width equal to the pump bandwidth (Eq. \ref{eq:filtered-joint-spectrum}),
and (c) filter width equal to $\frac{1}{5}$ of the pump bandwidth.\label{fig:JSI-theory}}

\vspace{0.2 in}

(a)\includegraphics[width=0.90in]{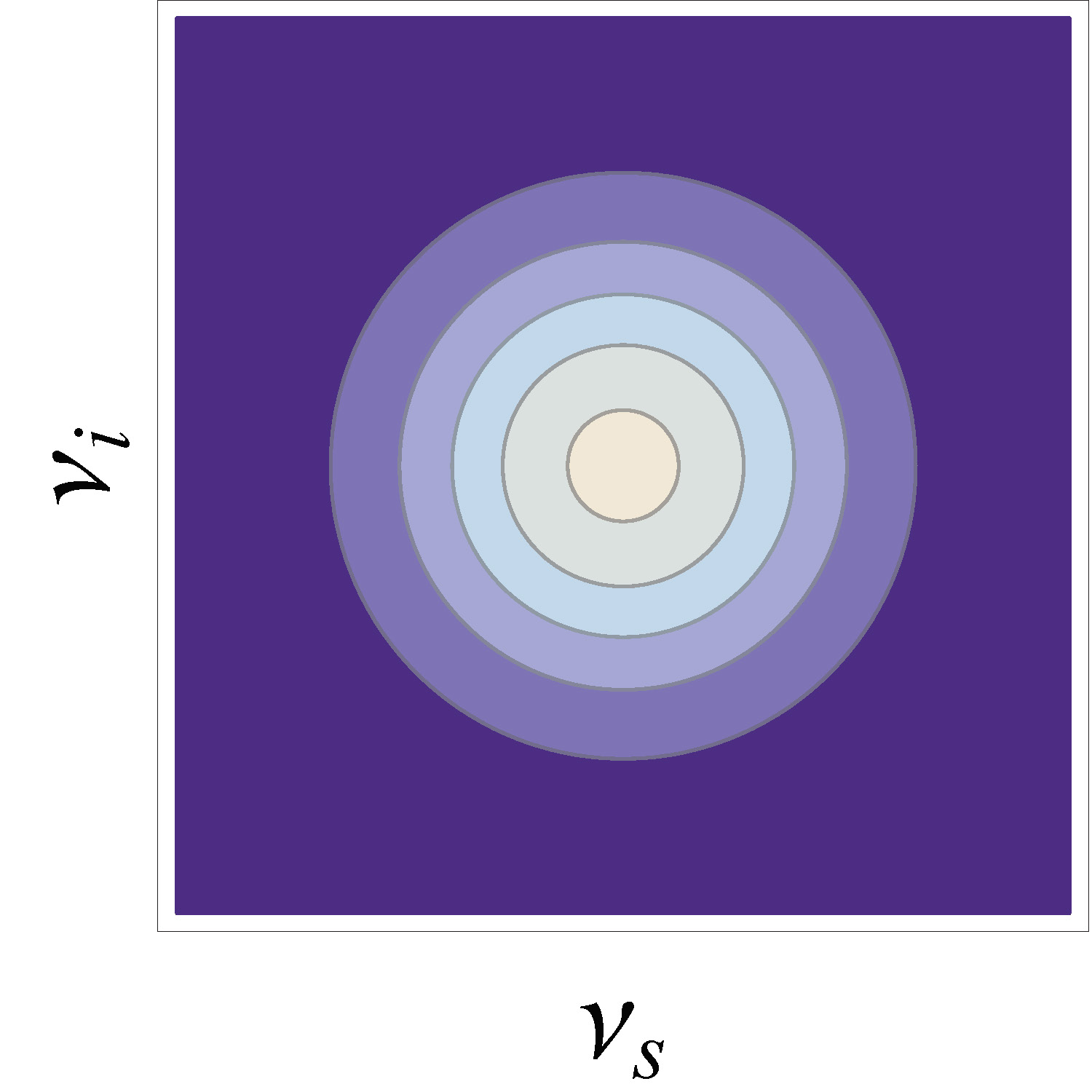}(b)\includegraphics[width=0.95in]{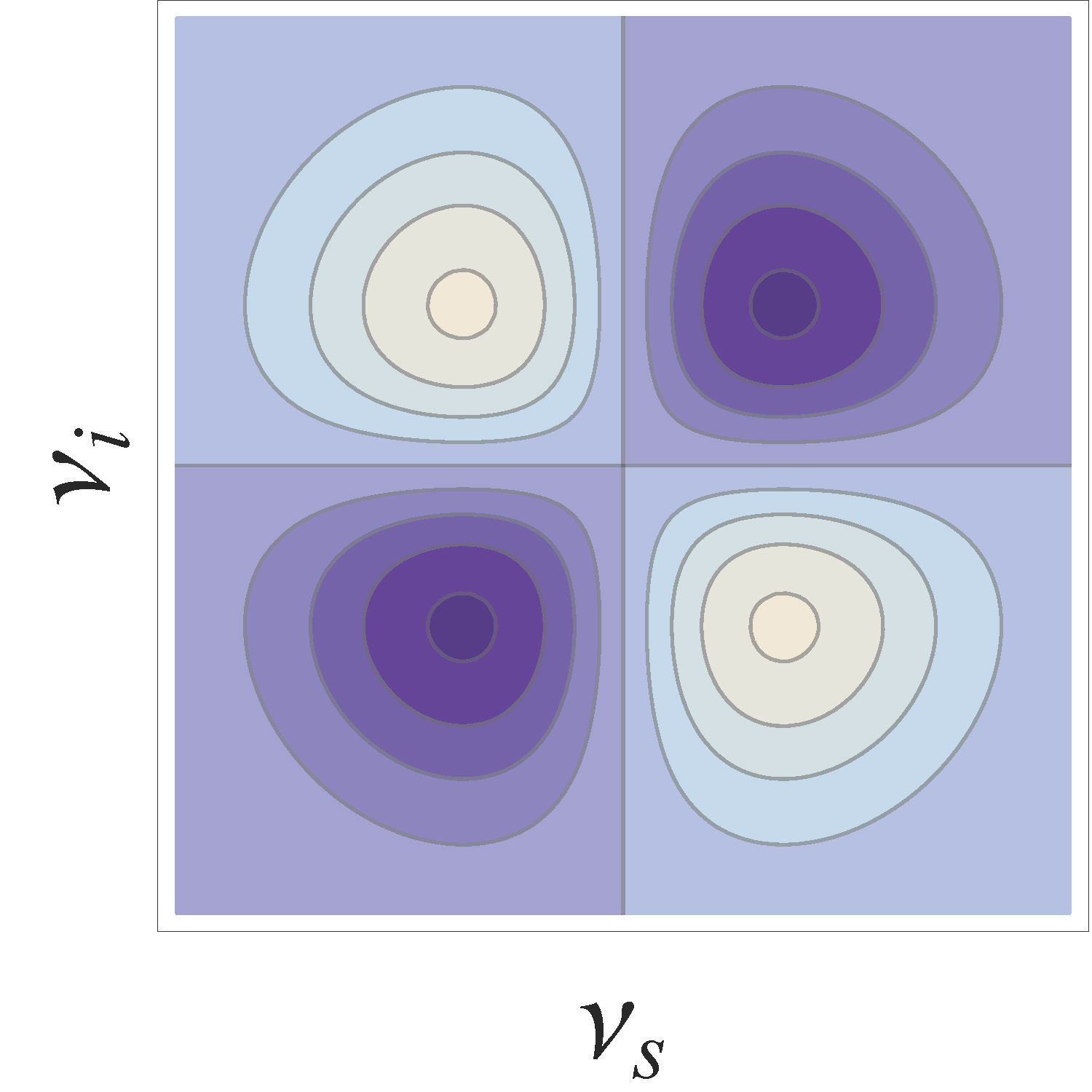}(c)\includegraphics[width=0.95in]{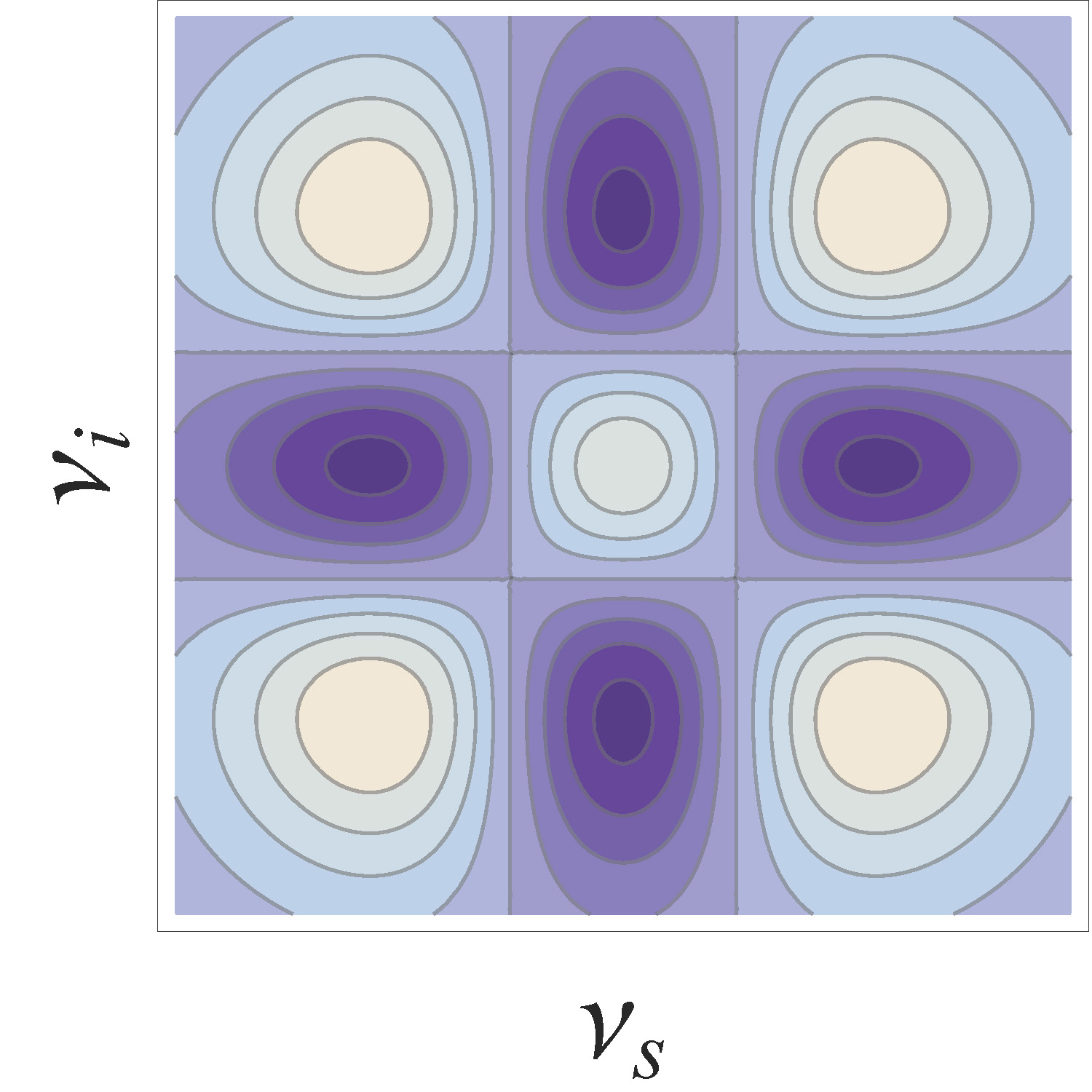}

\caption[Schmidt modes of the JSI]{(a-c) The first three Schmidt modes for a two-dimensional diagonal
Gaussian ellipse. Lighter areas are maxima; darker areas are minima.\label{fig:Schmidt-modes}}

\vspace{0.2 in}

(a)\includegraphics[width=0.90in]{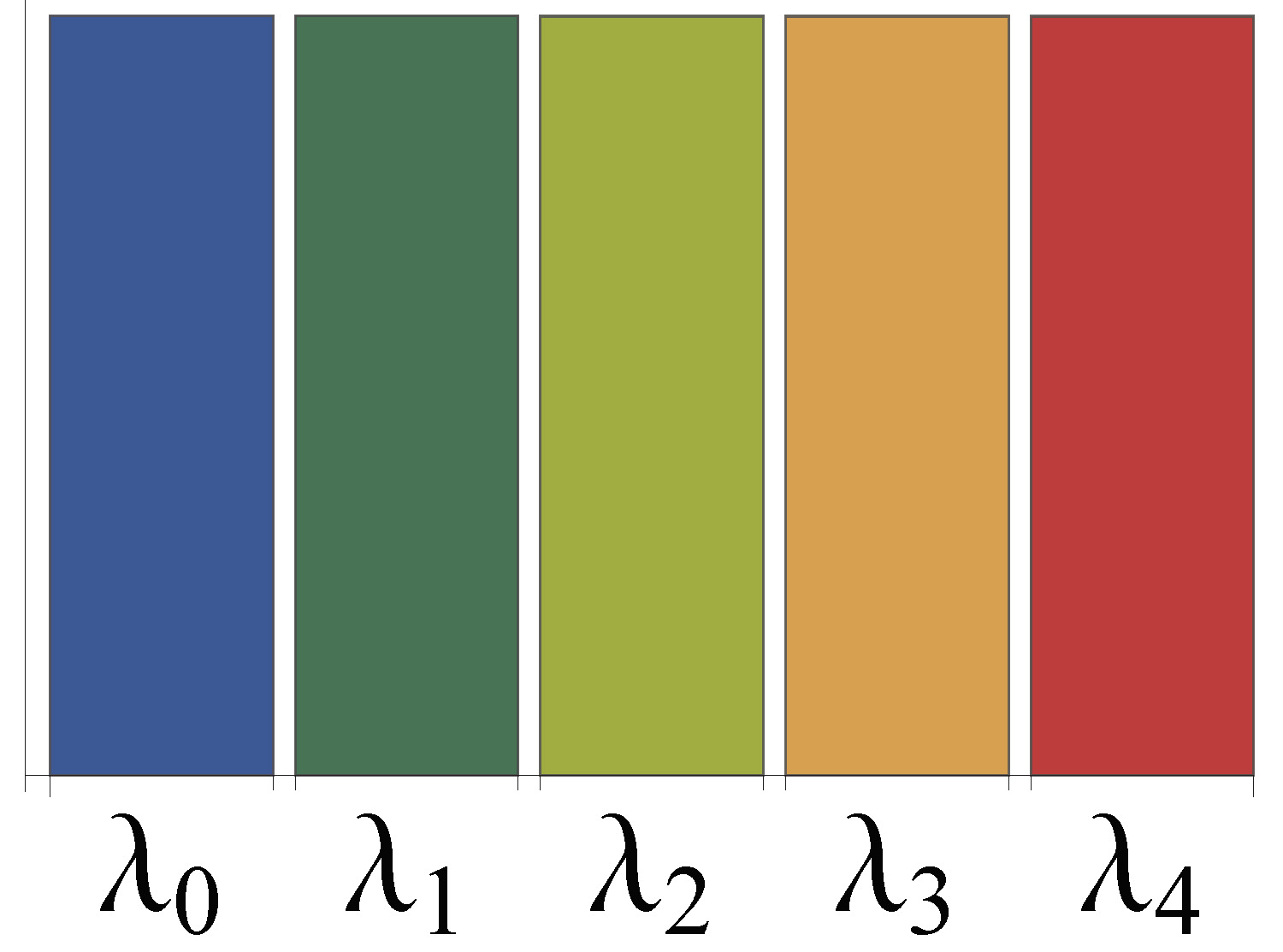}(b)\includegraphics[width=0.95in]{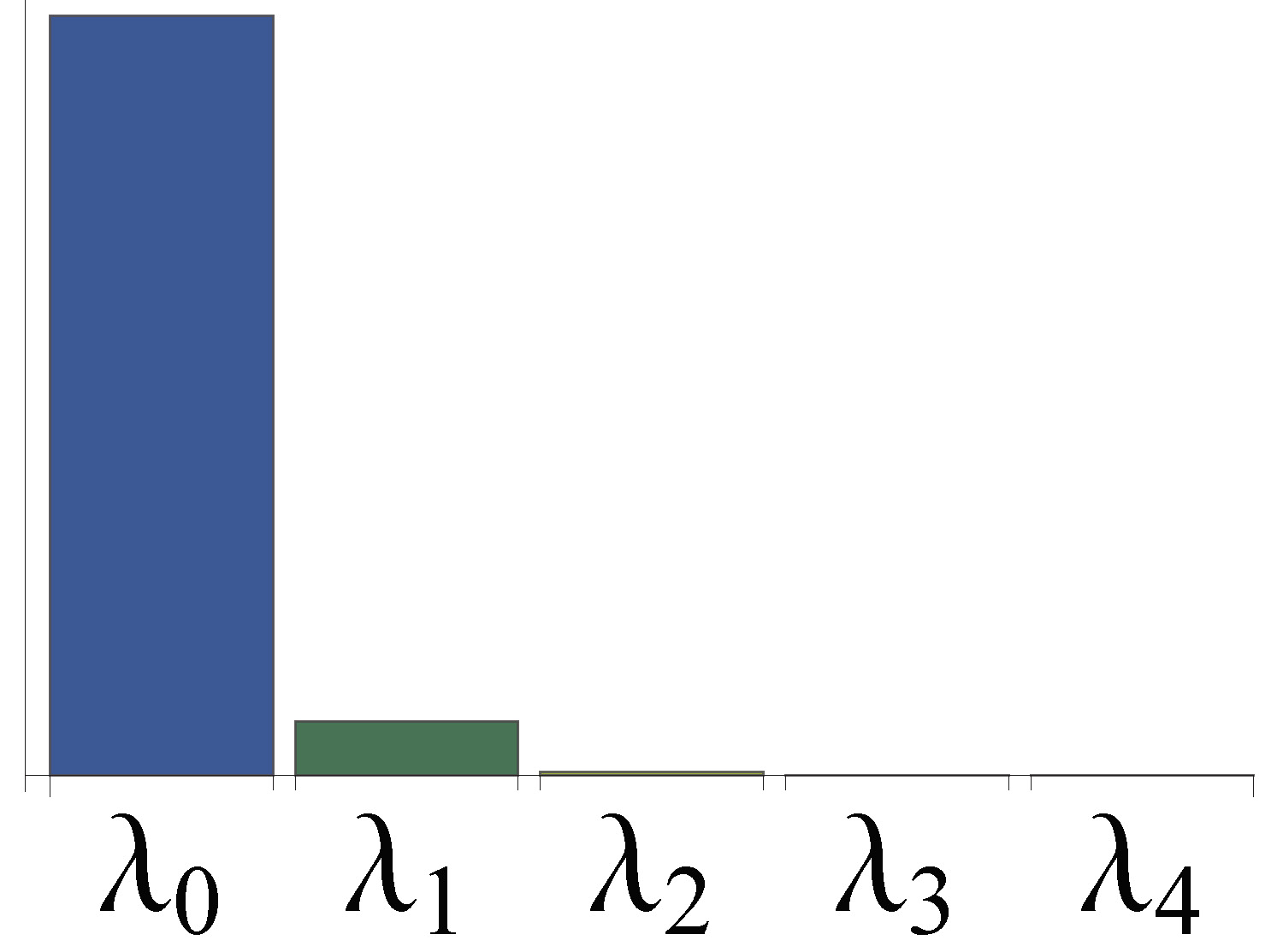}(c)\includegraphics[width=0.95in]{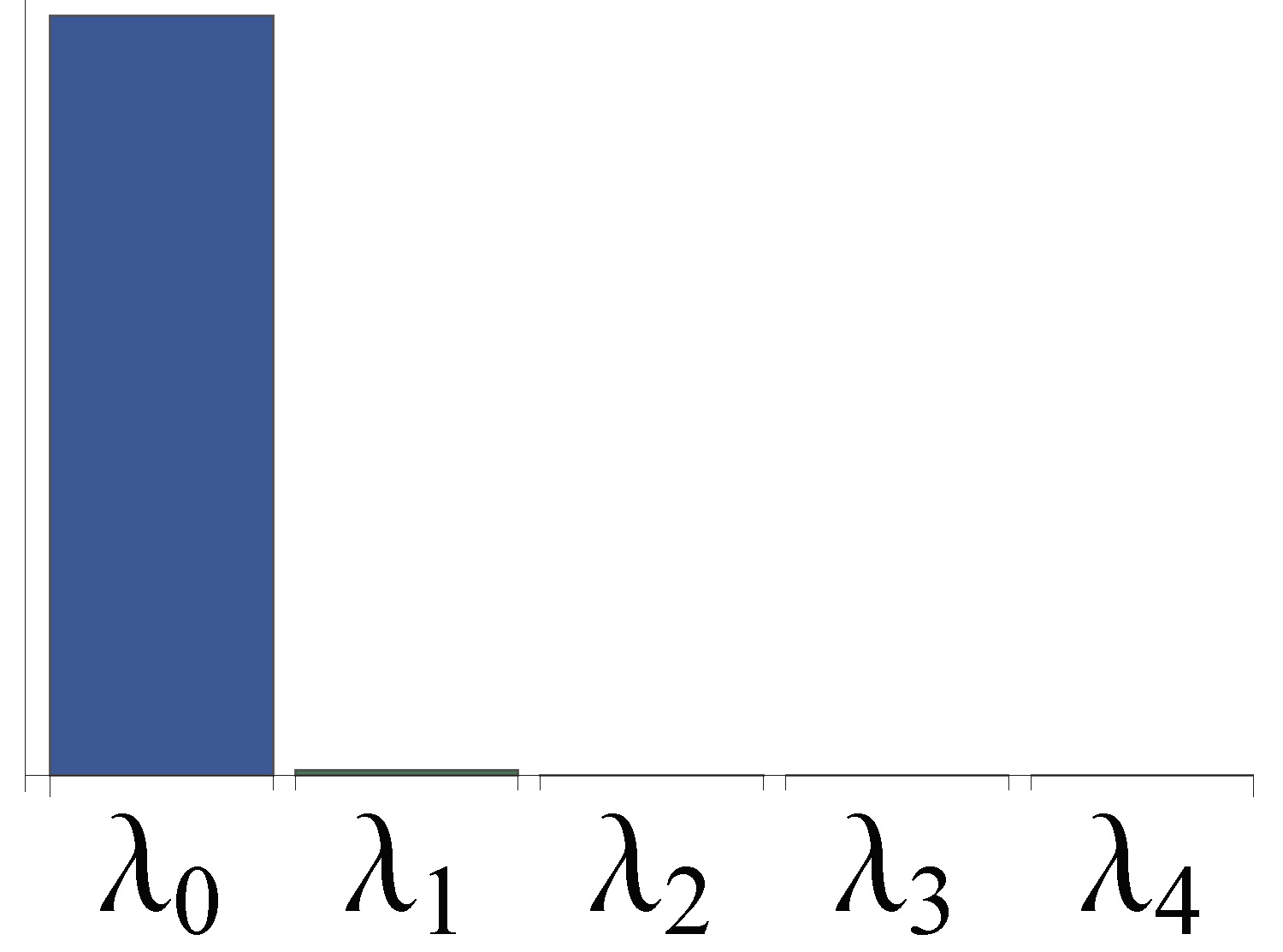}

\caption[Schmidt decomposition of modeled JSI]{Relative values of Schmidt decomposition eigenvalues for the unfiltered/filtered/highly-filtered
joint spectra of Fig.~\ref{fig:JSI-theory}. The resulting Schmidt
numbers (defined by Eq. \ref{eq:schmidt-number}) for the three cases
are (a) $K=\infty$, (b) $K=1.15$, and (c) $K=1.001$.\label{fig:Schmidt-values}}
\end{figure}

The joint spectral intensity, or joint spectrum,  $| f(\nu_{s},\nu_{i})| ^2$ can then be understood as the two-dimensional probability distribution associated with signal and idler emission frequencies, which we have plotted in  Fig.~\ref{fig:JSI-theory}(a) for a particular value of $\sigma$. In general, the joint spectrum may or may not be separable into functions that represent the spectral amplitudes of the signal and idler photons individually. The degree of non-separability determines how correlated the signal and idler photons are, which we will discuss in the next section.

\subsection{Schmidt Modes\label{sub:schmidt-modes}}

In order to quantify the degree of correlation between signal and idler photons one can apply the very useful Schmidt decomposition. The resulting Schmidt number can be used to quantify the entanglement between two systems, and hence the purity of each system individually. The Schmidt decomposition
\footnote{A practical note: calculating the Schmidt decomposition is functionally equivalent to singular value decomposition, which
is easily performed by any capable linear algebra package.} \cite{Nielsen2000,Eberly2006} allows a pure state of a composite
system $AB$ to be decomposed into a sum over products of orthonormal
states of $A$ and $B$:
\[
f_{A,B}=\sum_{i}\sqrt{\lambda_{i}}\,\,g_{A,i}\,\,g_{B,i},
\]
where the Schmidt coefficients $\sqrt{\lambda_{i}}$ satisfy the normalization
condition $\sum_{i}\lambda_{i}=1$. In a highly multi-dimensional space, this answers the difficult question of exactly which mode(s) to collect in order to maximize the probability of detection: the one(s) with the largest coefficients $\lambda_i$. The Schmidt coefficients can be used to define an
exceptionally useful quantity known as the Schmidt number%
\footnote{Confusingly, the Schmidt number is also sometimes defined as the total
number of eigenmodes \emph{without} weighting by eigenvalues, and $K$ is
sometimes referred to as the `cooperativity parameter.'%
} $K$, which can be described as the effective number of populated
eigenmodes:

\begin{equation}
K=\frac{1}{\sum_{i}\lambda_{i}^{2}}.\label{eq:schmidt-number}
\end{equation}

Why is this useful? First, it  naturally quantifies the degree
of entanglement in the physical system of interest, that also conveniently relates to the entropy
of the system, i.e.,  $\sum\lambda_i\log_{2}\lambda_i$ \cite{Law2000}.
Also, the inverse Schmidt number $1/K$ of the \emph{collected}
joint spectrum is equal to two other relevant quantities in an SPDC  or  SFWM source:
the purity $P$ of a heralded single photon and the visibility $V$
of a two-source HOM interferogram \cite{vicent_design_2010}. Note
that if $f_{A,B}$ is separable, the sum in Eq. \ref{eq:schmidt-number} is trivial, as by definition
we can simply write $f_{A,B}=g_{A}g_{B}$, so $K=1$. Similarly,
for a maximally polarization-entangled two-photon state $f_{A,B}=\left(H_{A}H_{B}+V_{A}V_{B}\right)/\sqrt{2}$,
$\lambda_{1} = \lambda_{2} = 1/2$ and the number of modes is $K=1/(\frac{1}{4}+\frac{1}{4})=2$. However,
in the case of Eq. \ref{eq:simple-joint-spectrum}, the continuous
basis and symmetry of the problem indicate that the Schmidt number
becomes a sum of $N\rightarrow\infty$ equally weighted terms, $\lambda_{i}=\frac{1}{N}$.
Thus, the Schmidt number is $K=1/(\sum_{i=1}^{N}1/N^{2})$, but the
denominator goes to zero as $N\rightarrow\infty$, indicating that
an infinite number of modes are required to describe this distribution.

Now suppose that we apply a Gaussian spectral filter (with bandwidth $\sigma_{f}$) to the signal
and idler modes. After filtering,
the joint spectrum from Eq. \ref{eq:simple-joint-spectrum} becomes


\begin{equation}
f(\nu_{s},\nu_{i})=A\exp\left[-\frac{\nu_{s}^{2}+\nu_{i}^{2}}{2\sigma^2}-\frac{2\nu_{s}\nu_{i}}{2\sigma^{2}}\right]
 \exp\left[-\frac{\nu_{s}^{2}}{2\sigma_{f}^{2}}\right]  \exp\left[-\frac{\nu_{i}^{2}}{2\sigma_{f}^{2}}\right],\label{eq:filtered-joint-spectrum}
\end{equation}

\noindent which is plotted in Fig.~\ref{fig:JSI-theory} for different
filter bandwidths. Note that if $\sigma_{f}$ is large, we recover
the unfiltered case of Eq. \ref{eq:simple-joint-spectrum}, while
if $\sigma_{f}$ is small, the filter term dominates and the joint
spectrum becomes separable. The Schmidt modes and their relative weights for varying filter bandwidths are shown in Figs.~2 and 3, respectively. In this general case, it can be shown
\cite{uren_photon_2003} that the inverse Schmidt number $1/K$ is
given by

\begin{equation}
\frac{1}{K}=\sqrt{1-\frac{1}{\left(1+(\frac{\sigma}{\sigma_{f}})^{2}\right)^{2}}},\label{eq:gaussian-schmidt-number}
\end{equation}

\noindent which has the expected behavior that $K\rightarrow\infty$ as $\sigma_{f}\rightarrow\infty$
(the unfiltered case), and $K\rightarrow1$ as $\sigma_{f}\rightarrow0$
(the tightly filtered case).    Note that while the discussion presented above, which is strictly valid only in the limit of a very short nonlinear medium,  is aimed at providing useful physical intuition, in a realistic situation the two-photon state is characterized by a joint spectrum which depends on phasematching properties as well as on the spatial shape of the pump, including the degree of focusing.  In general, the joint spectrum may be expressed as   $|f(\nu_s,\nu_i)|^2=|A(\nu_s,\nu_i)|^2 |\Phi(\nu_s,\nu_i)|^2$, where $\Phi(\nu_s,\nu_i)$ is determined by: i) the phasematching properties of the nonlinear medium (related to momentum conservation), and ii)  the \emph{spatial} shape of the pump field \cite{uren_photon_2003}.  For most situations, energy and momentum conservation lead to a joint spectrum  $|f(\nu_s,\nu_i)|^2$ which exhibits spectral correlations, so that spectral filtering can be used as discussed above in order to render the state factorable.

%
%

\subsection{The Spectrally Filtered Source}

As discussed in the introduction, an ideal photon pair source
for scalable optical quantum information processing would not exhibit joint spectral
correlations between the signal and idler photons. The discussion in the previous subsection
illustrates how these correlations arise from energy conservation,
and also how spectral filtering may eliminate them. However,  this solution has a significant drawback: if the quantum state
involves strong spectral correlations, the filters will block the great majority of the emitted
photon pairs. This can be seen easily from a plot of the joint spectrum
in Fig.~\ref{fig:JSI-theory}(a): an uncorrelated sub-ensemble of the emitted photon pairs
will lead to only a small fraction of the emitted photon pair-flux (in the theoretical limit of a perfectly correlated joint spectrum,
the fraction is actually zero). Another way to see this is through
the Schmidt decomposition: the best filtered, uncorrelated collection
mode one can hope for is the most populated Schmidt mode. If the Schmidt
number is large, there are many significantly populated Schmidt modes,
and any one of them will contain only a small fraction of the total
photon-pair flux.

Instead of employing spectral filtering to `fix' a highly correlated source, it would be desirable\footnote{It is worth noting that a filtered correlated source can actually be more desirable in some cases. For example, it is possible to take advantage of correlations to improve heralding efficiency (the chance of detecting an idler photon in a given mode conditional on the detection of a signal photon). However, this is achieved at the expense of reduced source brightness, i.e., fewer pairs overall.} to produce a joint spectrum which is already intrinsically uncorrelated. This is the motivation behind the various `engineered source' techniques \cite{Branning1999,grice_eliminating_2001,uren_photon_2003,cohen_2009,Smith2009,vicent_design_2010,
Rangarajan2011,Lutz2013,Fang2014dualpump,zielnicki2015engineering}.
We have characterized an engineered photon pair source that employs type-I degenerate
SPDC in $\beta$-barium borate (BBO), pumped with a pulsed 405 nm beam,
obtained from a frequency-doubled ultrashort  pulse train (35 fs time duration with 76-MHz repetition rate)
from a Titanium Sapphire laser centred at 810 nm; and photon-pair sources that employ SFWM in birefringent optical fibre, pumped with a pulsed 700 nm beam with 80 fs time duration and 80 MHz repetition rate from a Titanium Sapphire laser.

\section{Measuring the Joint Spectrum \label{chap:measuring-joint-spectrum}}

\subsection{Overview of Joint Spectral Measurement}

The most direct joint spectral measurement  possible would  register the signal and idler frequencies over many successive events, and from that estimate the underlying  probability distribution. However, directly measuring the frequency of a single photon is impractical, so we instead make use of
optical elements which map frequency to spatial mode.  For example, the angle at which light refracts from a prism or diffracts from an optical grating depends on frequency, allowing a measurement of the \emph{position} of a photon to determine its \emph{frequency}. This principle is used in a scanning monochromator, which uses a narrow slit  to determine the position (and thus the frequency) of a photon after a prism or grating. Thus, one could use two scanning monochromators counting photons in coincidence to construct a joint spectral intensity \cite{Kim2005,Mosley2008}, as described in Sec. \ref{sec:monochromator}. This technique is relatively simple and accurate, especially if using commercially available monochromators. However, it is not particularly fast, as any given photon pair is detected only if both photons pass through their monochromator slits, which for a reasonable resolution leads to a pair collection efficiency on the order of 0.1\%. One could upgrade the scanning monochromator technique to one which determines position with an array of, e.g., 20-40 single-photon counters, which must then be independently time-resolved or otherwise able to count in coincidence. Such a scheme in principle extracts usable information from every photon pair, but requires specialized detection and multicorrelation electronics \cite{Johnsen2014}.

Another possibility is to use Fourier transform spectroscopy to measure frequency in the time domain, as in Sec. \ref{sec:fourier-spectroscopy}. This exploits a self-interference effect and the property that the extremely short ({\raise.17ex\hbox{$\scriptstyle\sim$}}femtosecond) time regime of the electric field oscillations is easily accessible by an optical delay in a bulk optical medium. This provides a useful characterization of the joint spectrum, but is not an ideal measurement; it is not particularly simple due to the required scanning interferometers and the Fourier transform that relates the time-domain measurement to the desired frequency-domain result. It is also relatively slow, though Sec. \ref{sub:diagonal-fourier} describes a speed-up that makes the measurement more practical.

Dispersive-fibre spectroscopy, described in Sec. \ref{sec:Fibre-Spectroscopy}, provides another useful measurement of the joint spectral intensity \cite{Avenhaus2009}. In this scheme, we make use of dispersion in a long optical fibre to yield a frequency-dependent time of detection, from which the spectrum may be inferred.   This technique is one of the most direct measurements of the joint spectral intensity available, and is relatively fast and simple. The main disadvantage of dispersive-fibre spectroscopy is its high sensitivity to timing accuracy in detection electronics, and the lack of sufficiently low-loss dispersive fibres in all spectral regions.

Stimulated-emission-based measurement, described in Sec. \ref{sec:seb}, uses the corresponding stimulated version of the spontaneous photon-pair generation process to characterize the source \cite{Liscidini2013}. This measurement procedure requires an additional stimulating laser that is tunable across the signal or idler frequencies. The flux of stimulated photons is proportional to the flux of spontaneous photons with a proportionality constant equal to the number of photons in the stimulating seed, resulting in count rates many orders of magnitude higher than the spontaneously produced photons alone; thus, single-photon detectors are not required, only a standard spectrometer. This technique is one of the most efficient and high-resolution methods of measuring the joint spectral intensity, providing significantly shorter collection time and a significantly higher signal-to-noise ratio.

Another advantage of the stimulated-emission-based technique is that it can be extended to also measure the joint spectral \textit{phase}, as demonstrated in Ref. \cite{Jizan2016} (in this work we measure only the joint spectral intensity using this technique). The ability to measure the joint spectral phase is useful as correlations between the signal and idler photons can be present in the phase as well as in the intensity of the joint spectrum, for example when pumping the nonlinear medium with a non-transform-limited pulse train; note that if the pump is transform-limited, it is often acceptable to measure just the joint spectral intensity. In Sec.~\ref{chap:measuring-purity} we discuss two different measurement techniques which yield the heralded single-photon purity directly (rather than \emph{via} the joint spectrum) and which also include the effect of the joint spectral phase, namely $g^{(2)}$ correlation function measurement and two-source Hong-Ou-Mandel interferometry.

In this paper we present a comparison of these six measurement techniques, i.e., scanning monochromator measurement, Fourier spectroscopy, dispersive-fibre spectroscopy, stimulated-emission-based measurement, $g^{(2)}$ correlation function measurement, and two-crystal Hong-Ou-Mandel interferometry, specifically implemented for the sources described in Refs. \cite{zielnicki2015engineering} and \cite{Smith2009}. We demonstrate a subset of the techniques for each source. For the first source, which is based on frequency-degenerate and non-collinear photon pairs obtained via SPDC from a type-I BBO crystal pumped with  an ultrashort pump from a Ti:sapphire laser, we apply the techniques of Fourier spectroscopy, dispersive-fibre spectroscopy, $g^{(2)}$ correlation function measurement, and two-crystal Hong-Ou-Mandel interferometry. For the second source, which is based on non-degenerate photon-pairs obtained via SFWM from a polarization-maintaining optical fibre pumped with an ultrashort pump from a Ti:sapphire laser (degenerate pumps), we apply the techniques of scanning monochromator measurement, stimulated-emission-based measurement, and $g^{(2)}$ correlation function measurement. In the conclusion we compare all of these techniques' relative advantages and disadvantages and quantify their resolution, sensitivity, and efficiency.

\subsection{Scanning Monochromator Technique}\label{sec:monochromator}

In the scanning monochromator measurement, shown in Fig.~\ref{fig:monochromatorsetup}, the signal and idler photons are directed into two separate diffraction grating-based monochromators.     At the output ports of the monochromators,  the photons are coupled into a multi-mode fibre, and a frequency scan is performed by rotating the gratings: only the frequency components coupled into the fibres are recorded and thus the joint spectrum can be collected as a series of pairs of frequency values. The photon pairs are detected with avalanche photodiodes and counted in coincidence. A scan is performed over the dual frequency range by holding one diffraction grating at a constant orientation while rotating the other diffraction grating, repeating this procedure for a vector of angular orientation values for the first grating, thus performing a two-dimensional sweep of the signal and idler frequencies. The rate at which coincidences are observed is very low, and thus each data point requires a long integration time, limited by the stability of the source over the time required for the overall procedure. Thus, while the procedure is conceptually straightforward, it is clearly inefficient.

\subsubsection{Scanning Monochromator Measurement for the SFWM Source}\label{sec:monochromatormeas}

\begin{figure}
\begin{center}
\includegraphics[width=3in]{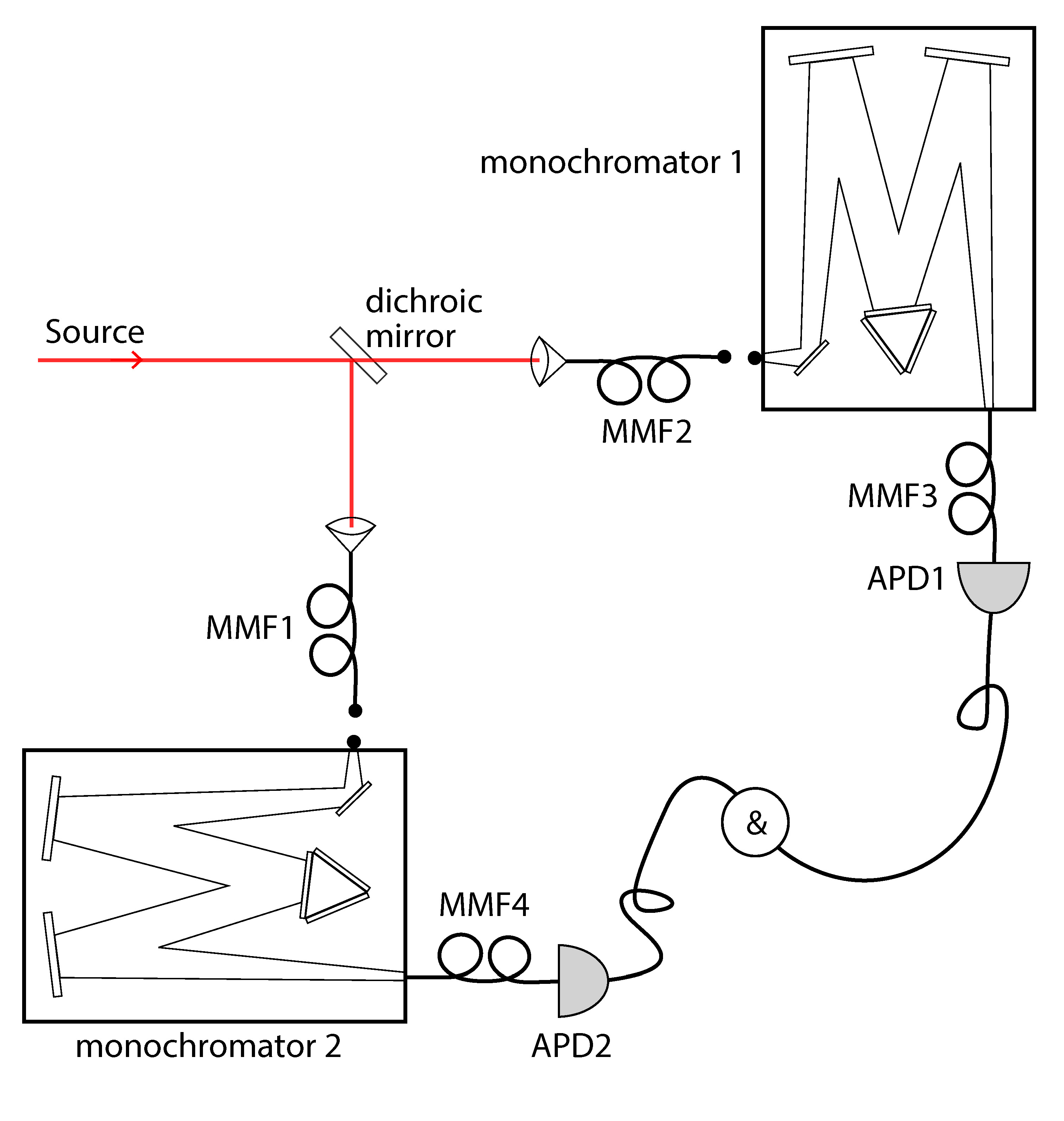}
\end{center}
\caption[Scanning Monochromator Setup]{\label{fig:monochromatorsetup}
Schematic of the scanning monchromator technique. MMF: multi-mode fibre, APD: avalanche photodiode.}
\end{figure}

\begin{figure}
\begin{center}
\includegraphics[width=3.2in]{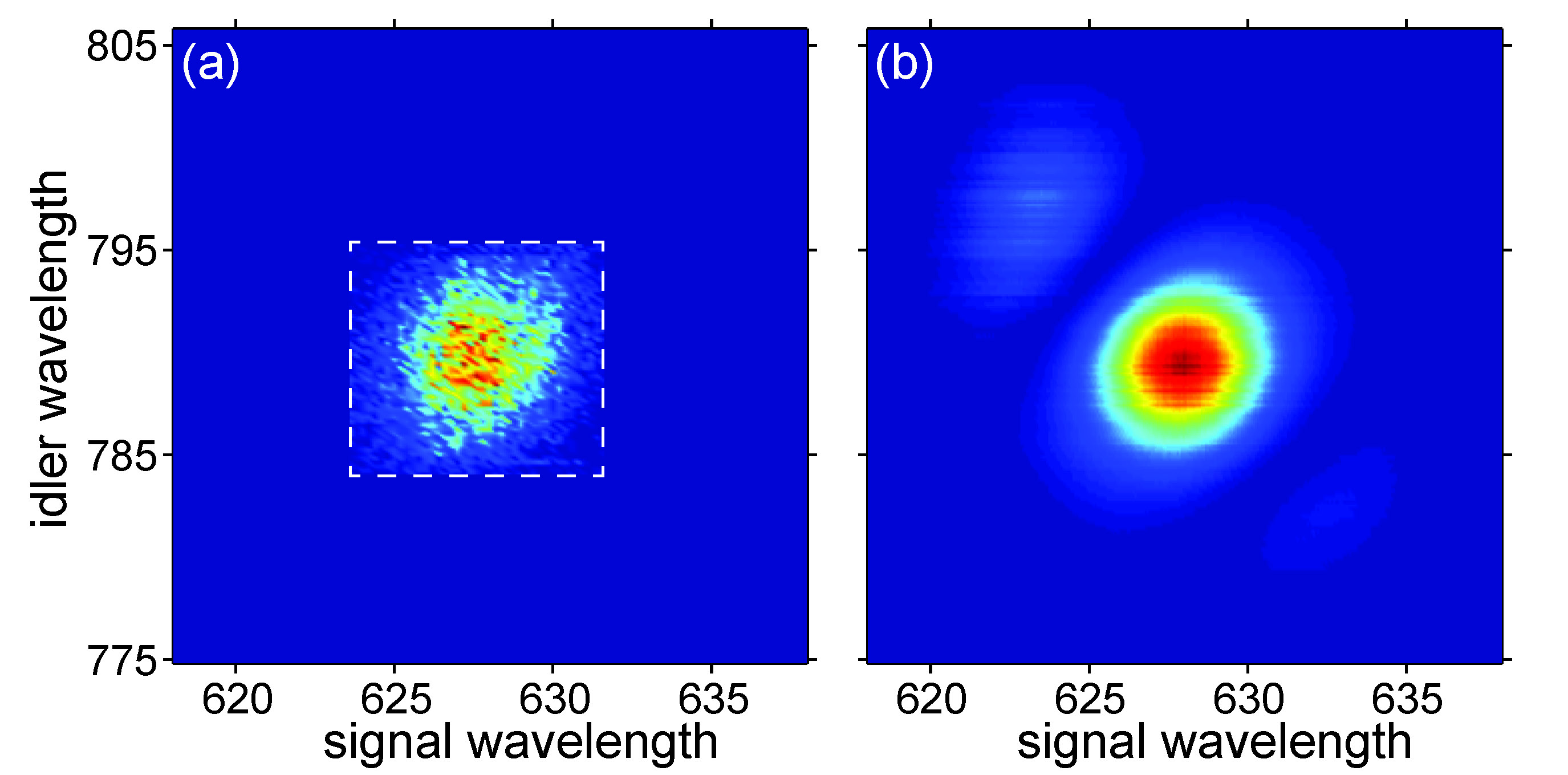}
\end{center}
\caption[Scanning Monochromator Measurement]{\label{fig:fibremeas}
Joint spectral intensity of photon pairs produced in a bow-tie fibre SFWM source, (a) from the scanning monochromator measurement, (b) from the stimulated-emission-based measurement.}
\end{figure}

Figure \ref{fig:monochromatorsetup} shows a schematic of the experimental setup for the scanning monochromator measurement. We apply the technique to a bow-tie polarization-maintaining optical-fibre SFWM source \cite{garay2016photon,cruz2016fiber}. We use an ID Quantique 800 time-tagger module to count coincidences, with the coincidence window set to be 5.67 ns; the repetition rate of the laser is 80 MHz, so that this window is less than half of the time between pulses. We scan over the dual frequency range by holding one diffraction grating at a constant orientation and taking 0.2 nm spectral steps across the other diffraction grating, covering the range 624-632 nm for the signal and 784-796 nm for the idler. The rate at which we observe coincidences is very low, so we integrate for 60 s in each grating position; the whole procedure takes approximately 40 hours. The results are shown in Fig.~\ref{fig:fibremeas}(a); the dotted square represents the spectral area where data were taken.   While the presence of a peak is clear, there are very few counts for each point, resulting in a low signal-to-noise ratio. The number of counts would increase with an even longer integration time, but the increase of signal over noise must be balanced with the effects of drift over the time required for the overall procedure. Note that the purity, calculated from the measured joint spectral intensity while ignoring any possible joint phase effects, of the heralded single-photon state is $0.826\pm0.004$.


\subsection{Fourier Spectroscopy\label{sec:fourier-spectroscopy}}

\begin{figure}
\begin{center}
(a)\includegraphics[scale=0.18]{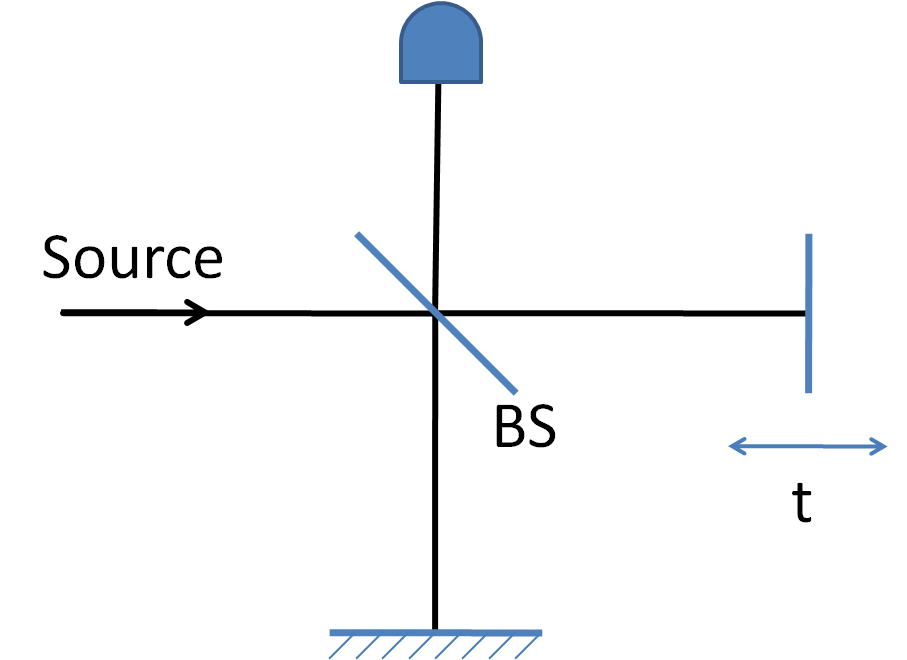}
(b)\includegraphics[scale=0.18]{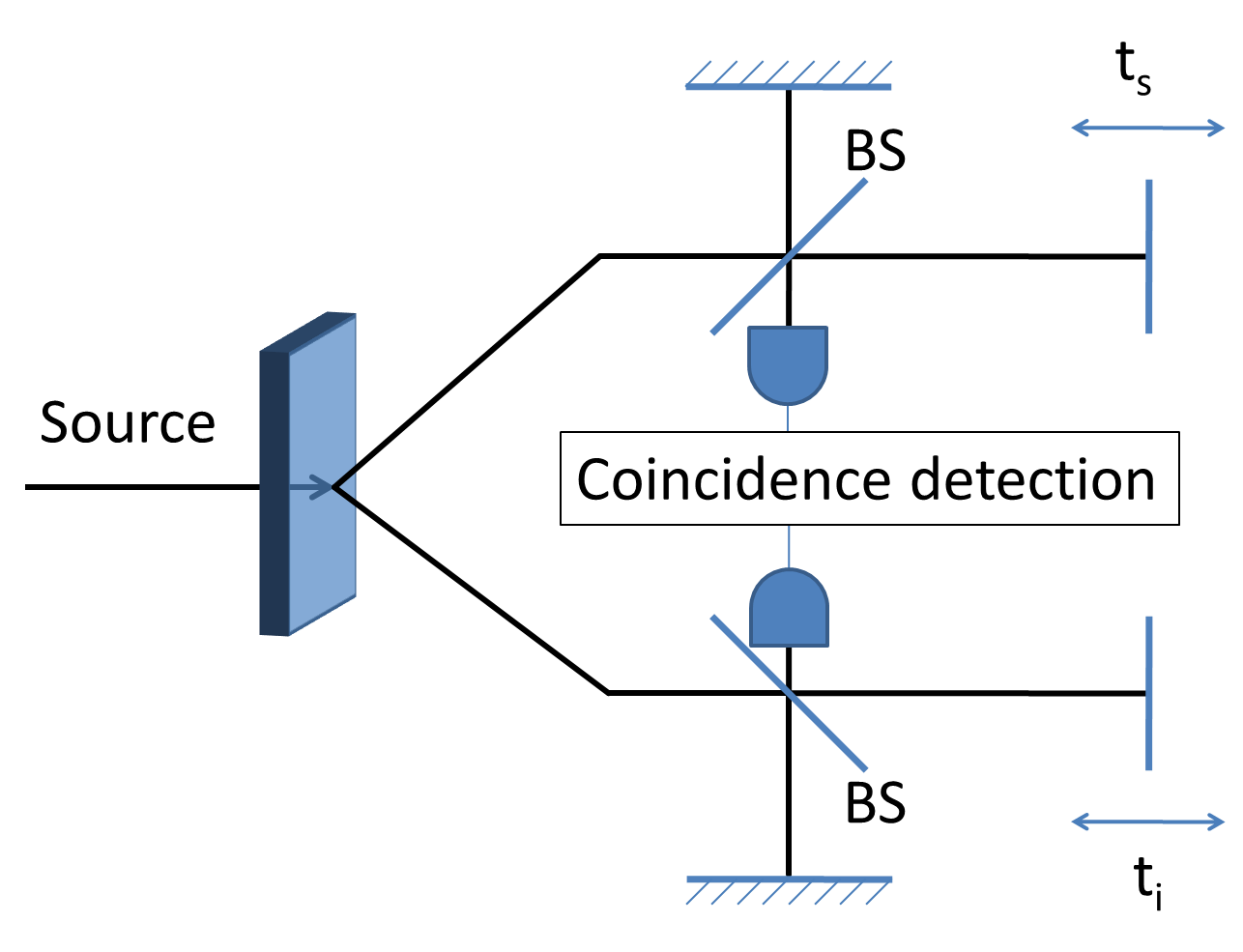}
\end{center}

\caption[Fourier spectroscopy with Michelson interferometers]{\label{fig:Fourier-spectroscopy}
Schematic diagram of (a) one-dimensional and (b) two-dimensional Fourier spectroscopy. Both systems use Michelson interferometers, but in the two-dimensional case, the interferometer is duplicated in both arms of an SPDC source, and counts are measured in coincidence.}
\end{figure}

The well-established technique of Fourier
spectroscopy \cite{hariharan_optical_2003} relies
on the fact that a Michelson interferometer with a variable relative path length in one of the two arms
can be used to extract the spectrum of the incoming light from the Fourier transform of a time-domain interferogram.
Light with an unknown spectrum is sent into a Michelson interferometer, as shown in Fig.~\ref{fig:Fourier-spectroscopy}(a).
Recording the output intensity while scanning the path length difference in the interferometer,  a time-domain signal
is measured from which the spectrum can be obtained through a Fourier transform followed by appropriate numerically-implemented spectral
filtering.   Mathematically, the time-domain interferogram $\tilde{I}(\tau)$ is related to the incoming spectral intensity $ I(\omega)$
through a Fourier cosine transform, with an offset term, as follows

\[
\tilde{I}(\tau)\varpropto\int_{0}^{\infty}d\omega I(\omega)[1+\cos(\omega\tau)],
\]

So, by simply computing the inverse transform on the measured data, the spectrum
of the signal is recovered:
\[
I(\omega)\varpropto\int_{0}^{\infty}d\tau\left(I(\tau)-\frac{1}{2}I(\tau=0)\right)\cos(\omega\tau).
\]
\subsubsection{Two-Dimensional Fourier Spectroscopy\label{sec:2D-fourier}}

\begin{figure}
\begin{centering}
\includegraphics[width=3.2in]{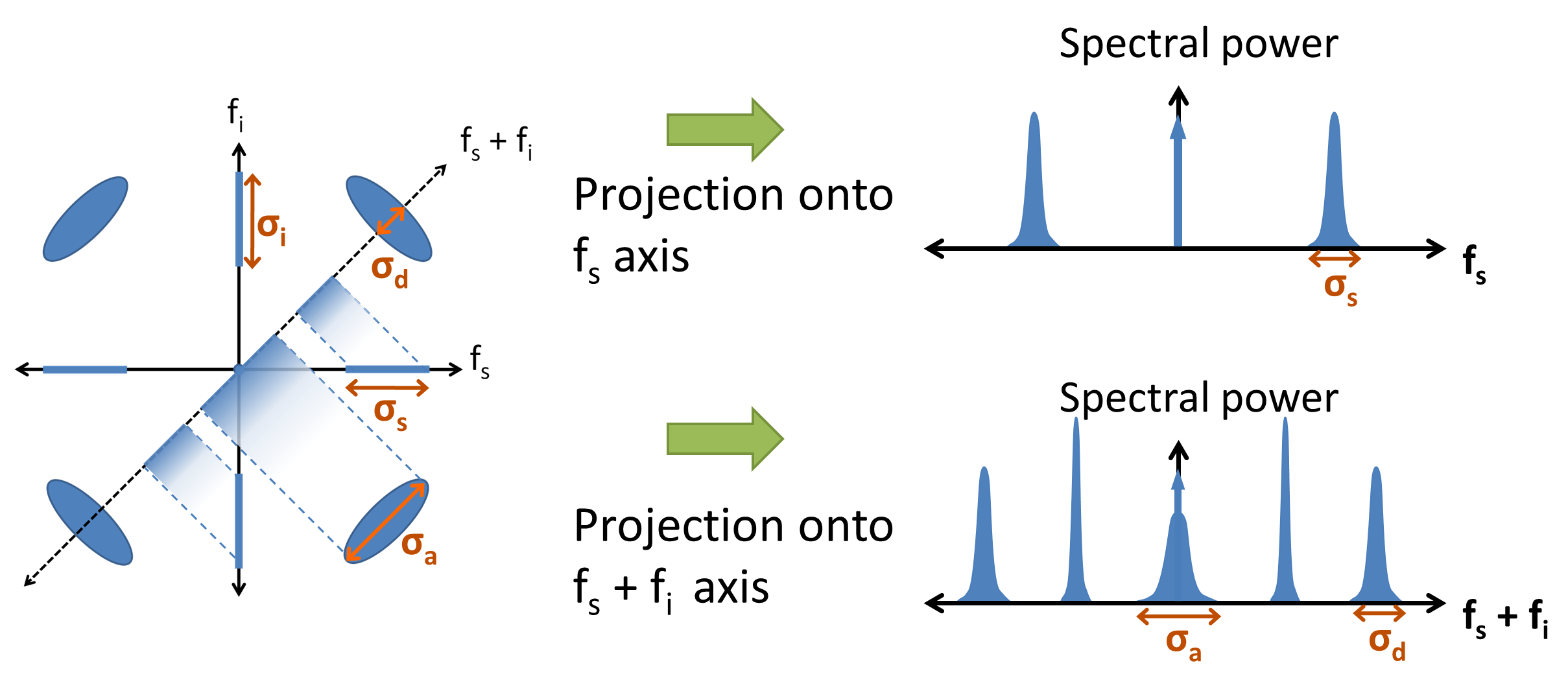}
\par\end{centering}

\caption[Diagonal Fourier spectroscopy]{\label{fig:Frequency-domain-picture}Sketch of the frequency-domain
signal resulting from two-dimensional Fourier spectroscopy.}
\end{figure}

In the context of characterizing an SPDC photon-pair source, performing Fourier
spectroscopy on either the signal or idler arm permits measuring
the corresponding single-photon spectrum.  In order to measure the \emph{joint} spectrum, we employ a generalized version of this technique \cite{wasilewski_joint_2006} in which scanning interferometers are placed in both the signal
and idler arms, while collecting coincidence counts, as shown in
Fig.~\ref{fig:Fourier-spectroscopy}(b). The time-domain data collected
by independently scanning the interferometers are then related to the
Fourier transform of the joint spectrum:


\begin{equation}
\tilde{I}(\tau_{s},\tau_{i})\varpropto\int_{0}^{\infty}d\omega_{s}\int_{0}^{\infty}d\omega_{i}I(\omega_{s},\omega_{i})
\left(1+\cos(\omega_{s}\tau_{s})\right)\left(1+\cos(\omega_{i}\tau_{i})\right),
\end{equation}

\noindent where $\tilde{I}(\tau_{s},\tau_{i})$ is the joint temporal intensity (JTI)
and $I(\omega_{s},\omega_{i})$ is the joint spectral intensity (JSI).
Analogously to the one-dimensional case, performing a 2D Fourier transform
on the measured data and retaining only terms in the $\omega_s>0,\omega_i>0$ quadrant gives

\begin{multline}
\int d\tau_{s}d\tau_{i}\tilde{I}(\tau_{s},\tau_{i}) \exp(i\omega_{s}\tau_{s}+i\omega_{i}\tau_{i})\\
\varpropto\delta(\omega_{s})\delta(\omega_{i})\left\langle \hat{N}_{s}\hat{N}_{i}\right\rangle +\frac{1}{2}\delta(\omega_{s})\left\langle \hat{N}_{s}\hat{I}_{i}(\omega_{i})\right\rangle +\\
\frac{1}{2}\delta(\omega_{i})\left\langle \hat{N}_{i}\hat{I}_{s}(\omega_{s})\right\rangle +\frac{1}{4}I(\omega_{s},\omega_{i}),
\end{multline}

\noindent where $\hat{I}_{i}(\omega_{i})$ and $\hat{I}_{s}(\omega_{s})$
represent the spectral intensities of the signal and idler beams, and $\hat{N}_{\mu}=\int d \omega  \hat{I}_{\mu}(\omega) $  (with $\mu=s,i$) are the total number operators for
signal (s) and idler (i) photons.  While the terms shown represent the top-right quadrant, as shown in Fig.~\ref{fig:Frequency-domain-picture}, symmetric terms also appear in the other three quadrants. The first term is
located at the origin and is proportional to the total number of coincidence
counts. The middle terms are located on the axes and provide information
about the single-photon spectrum of the signal and idler photons, conditioned on the
detection of the conjugate photon. The term of interest is the final
term, which is proportional to the JSI.

\subsubsection{Diagonal Fourier Spectroscopy\label{sub:diagonal-fourier}}

Unfortunately, collecting two-dimensional data is very time consuming
compared to one-dimensional Fourier spectroscopy, requiring $N^{2}$
rather than $N$ points to obtain the same resolution. However, under the assumption that the JSI is approximately Gaussian (which is valid for our source), we
can take advantage of the structure of the two-dimensional spectrum
to measure the relevant parameters with a one-dimensional scan. Specifically, because the JSI is well approximated by a Gaussian ellipse with its major and minor axes aligned with the diagonal frequency axes $\omega_s+\omega_i$ and $\omega_s-\omega_i$, the spectral correlations can be well characterized by the diagonal widths along these two axes.  Thus, the Fourier transform of a 1D scan along the $t_{s}+t_{i}$  ($t_{s}-t_{i}$) axis yields the projection of the 2D spectrum along the $\omega_{s}+\omega_{i}$ ($\omega_{s}-\omega_{i}$) axis,  as shown in Fig.~\ref{fig:Frequency-domain-picture}.
If we model the joint frequency spectrum as a 2D Gaussian ellipse,
the relevant parameter for spectral correlation is the ratio of the peak widths along the
$\omega_{s}+\omega_{i}$ and $\omega_{s}-\omega_{i}$ axes, $\sigma_d/\sigma_a$.
The parameters $\sigma_{d}$ and $\sigma_{a}$ can be extracted directly
from the 1D scan described above, characterizing the Gaussian ellipse
\begin{equation}
f(\nu_{s},\nu_{i})=A\exp \left[-\left(\nu_{s}^{2}+\nu_{i}^{2}\right)\left(\frac{1}{4\sigma_{d}^{2}}+\frac{1}{4\sigma_{a}^{2}}\right)
-2\nu_{s}\nu_{i}\left(\frac{1}{4\sigma_{a}^{2}}-\frac{1}{4\sigma_{d}^{2}}\right)\right].\label{eq:diagonal-ellipse}
\end{equation}
Recall from Eqs. \ref{eq:filtered-joint-spectrum} and \ref{eq:gaussian-schmidt-number} that we can determine the heralded single-photon purity $P$ (or equivalently,
the inverse Schmidt number $1/K$) directly from a Gaussian ellipse.
Applying a change of variables and solving for $\sigma_{a}$ and $\sigma_{d}$ in terms
of $\sigma$ and $\sigma_{f}$, we can rewrite Eq. \ref{eq:gaussian-schmidt-number}
as
\begin{equation}
P=\sqrt{1-\left(\frac{r-1}{r+1}\right)^2},
\label{eq:heralded-purity2}
\end{equation}
where $r\equiv\sigma_{d}^{2}/\sigma_{a}^{2}$.

\subsubsection{Fourier Spectroscopy Measurements for the SPDC Source\label{sec:fourier-meas}}

\begin{figure}
\begin{flushleft}
(a)\quad\includegraphics[width=1.8in]{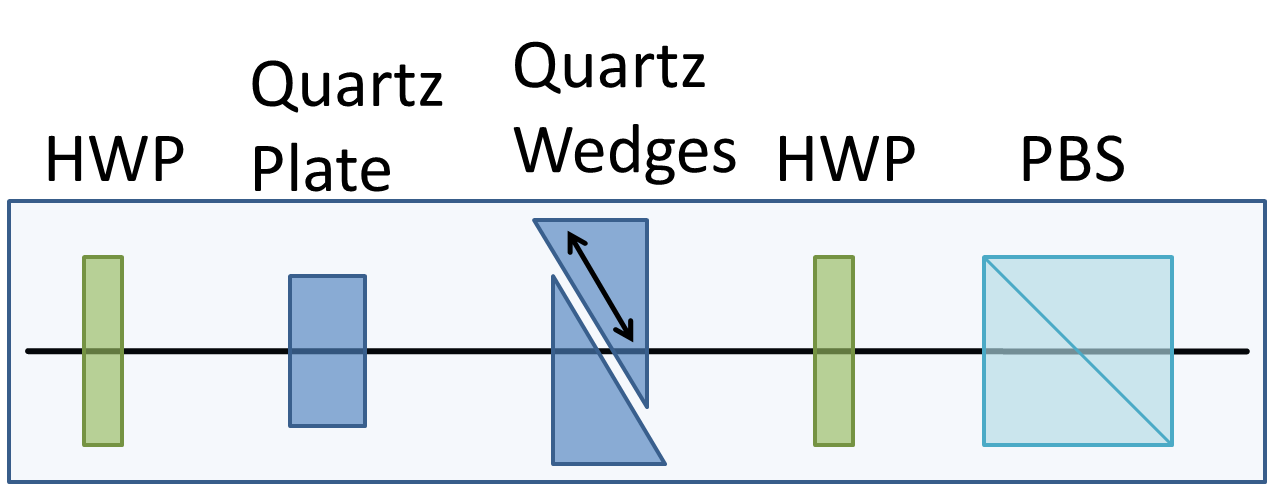}
(b)\includegraphics[width=3.2in]{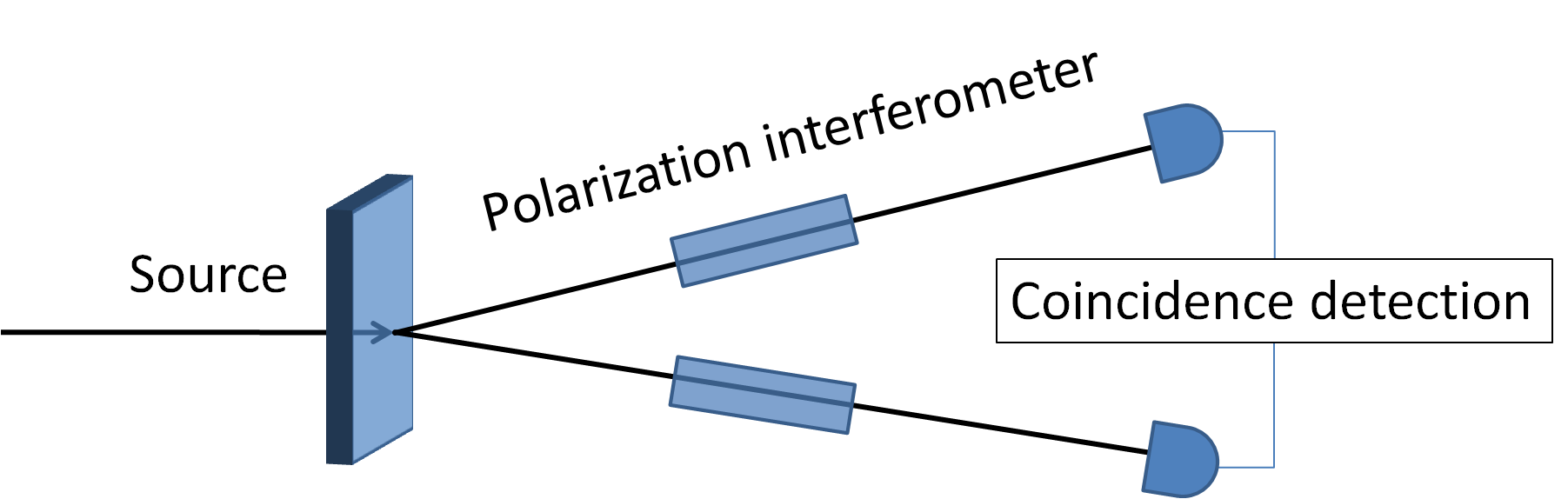}
\end{flushleft}

\caption[Fourier spectroscopy schematic]{\label{fig:Fourier-spectroscopy-setup}Diagram of 2D Fourier spectroscopy
setup for measuring the joint spectrum. The common-path polarization interferometer
in (a) uses a half-wave plate (HWP) to rotate light into the diagonal
basis, followed by a birefringent quartz plate to initially delay
horizontally polarized light (H) relative to vertically polarized
light (V). Then, quartz wedges are used to variably delay V relative
to H. Finally, another HWP rotates back into the H/V basis and a polarizing
beam splitter (PBS) is used to pick off the H component. Two of these
polarization interferometers are used in (b) to analyse the joint
spectrum.}
\end{figure}

Figure \ref{fig:Fourier-spectroscopy-setup} shows our experimental setup for the Fourier spectroscopy measurements. The critical difference between this and the theoretical discussion above is the use of a common-path polarization interferometer rather than a Michelson. This achieves the same function by providing an optical path length difference due to the different indices of refraction of ordinarily and extraordinarily polarized light in birefringent quartz, rather than a physical path length difference. The advantage of this technique is that the two optical paths take the same physical path, making the interferometer much more robust against vibrations and thermal fluctuations. Specifically, horizontally polarized light is rotated into the diagonal basis using a half-wave plate\footnote{Due to the broad bandwidth involved, one must take care to use waveplates with a flat retardance (that is, the optical path-length difference between ordinary and extraordinary polarization $L_e-L_o$) in the wavelength region of interest.} before passing through scanning quartz wedges with vertical optic axes. Diagonally polarized light is a superposition of ordinary (o) and extraordinary (e) polarization in the crystal, which pick up different phases,
\[
\left|D\right\rangle = \frac{\left|e\right\rangle + \left|o\right\rangle}{\sqrt{2}} \rightarrow \frac{1}{\sqrt{2}} \left( e^{2 \pi i n_e L / \lambda}\left|e\right\rangle + e^{2 \pi i n_o L / \lambda}\left|o\right\rangle \right),
\]where $L$ is the length of quartz, $\lambda$ is the wavelength and $n_o$ and $n_e$ are the ordinary and extraordinary indices of refraction at that wavelength. For our quartz wedges (custom-made by Rocky Mountain Instruments) at 810 nm, with a wedge angle of $32^\circ$ and one wedge mounted on a translation stage, this corresponds to a relative delay of 15.75 fs per mm of stage motion (where 2.7 fs corresponds to a relative phase of $2\pi$). A fixed-length quartz plate is used to provide a delay offset in the opposite direction using an optic axis mounted orthogonally to that of the wedges. Finally, the polarization is rotated back to the horizontal (H)/ vertical (V) basis and analysed with a polarizing beamsplitter. The probability $P$ of observing a horizontally polarized photon, i.e., the photon exiting the horizontal port of the PBS, depends on the relative phase $\Delta\phi \equiv 2\pi L (n_o - n_e) / \lambda$ as
\[
P = \cos(\Delta\phi)^2 = \frac{1}{2}\left(1+\cos(2\Delta\phi)\right).
\]

\begin{figure}

\begin{centering}
\includegraphics[width=3.1in]{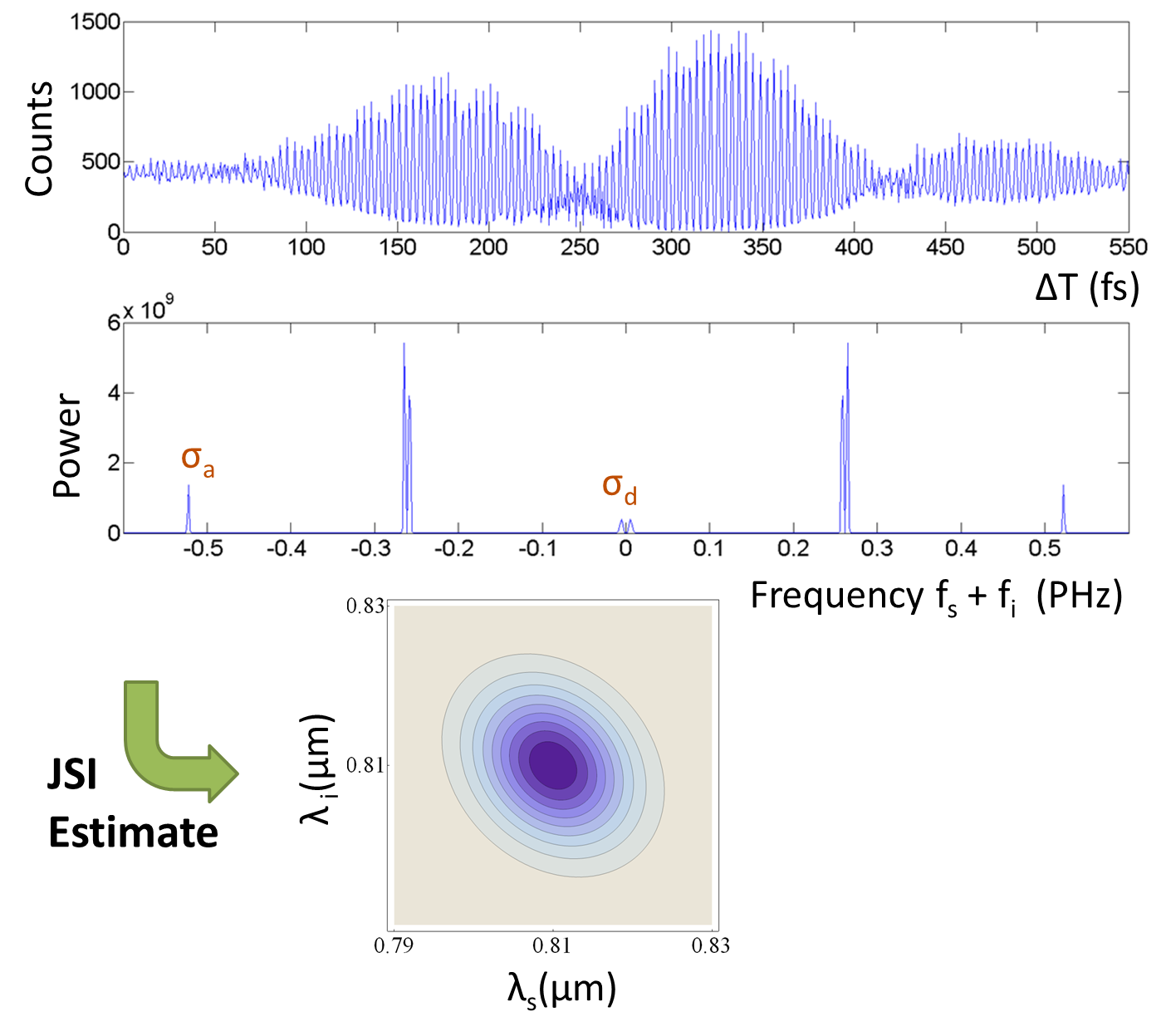}
\par\end{centering}

\caption[Measured joint spectrum from Fourier spectroscopy]{\label{fig:Experimental-joint-spectrum} Experimental measurement of
joint spectrum for a pump bandwidth of 5.2 nm, collected through 20 nm filters, resulting
in an estimated purity of $0.96\pm0.02$. The ellipse represents the estimated joint spectrum,
with the eccentricity along the diagonal providing spectral correlation
(or in this case, anti-correlation). The purity here applies
only to the spectral state of heralded single photons, and represents
the degree of factorability of the joint spectral intensity, i.e., not including any possible phase correlations. The double peaks seen in the frequency plot are due to a slight misalignment of the diagonal scan axis and are fit with two identical offset Gaussian peaks.}
\end{figure}

Fig.~\ref{fig:Experimental-joint-spectrum} shows typical
results from applying this technique to our source, and illustrates the reconstruction of the joint spectrum from the measured data. The spectrum is Fourier transformed to the frequency domain, and then a Gaussian fit is applied to peaks corresponding to the diagonal widths of the joint spectrum. From this, the heralded single-photon purity can be determined through Eq. \ref{eq:heralded-purity2}.

\begin{figure}
\begin{centering}
\includegraphics[width=3.0in]{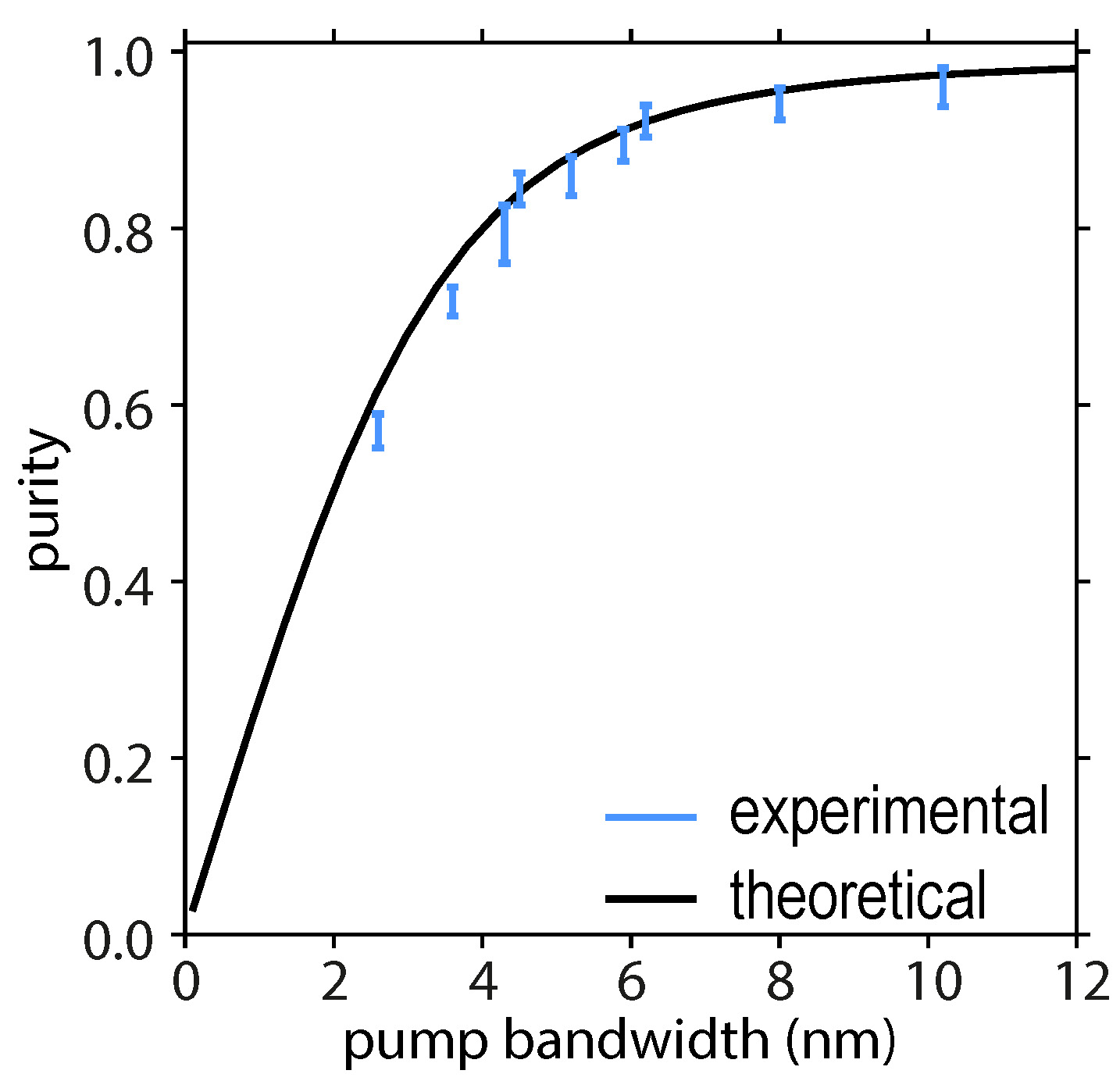}

\par\end{centering}

\caption[Single-photon purity]{\label{fig:purity}Single-photon purity of our SPDC source, as measured with our diagonal Fourier transform spectroscopy setup, as a function of the pump bandwidth.   As expected, decreasing the pump bandwidth has the effect of increasing the photon-pair degree of spectral entanglement, thus increasing the Schmidt number $K$ and decreasing the single-photon purity $P=1/K$.}

\end{figure}

As discussed above, the joint spectrum is given in general as the product  $|f(\nu_s,\nu_i)|^2=|A(\nu_s,\nu_i)|^2 |\Phi(\nu_s,\nu_i)|^2$.   Note that for a large pump bandwidth the function $|\Phi(\nu_s,\nu_i)|^2$ will tend to dominate over a comparatively broad $|A(\nu_s,\nu_i)|^2$ so as to determine the resulting shape of the joint spectrum $|f(\nu_s,\nu_i)|^2$.  Conversely, in the limit of a monochromatic pump $|A(\nu_s,\nu_i)|^2 \rightarrow \delta(\nu_s+\nu_i)$, the pump envelope function $|A(\nu_s,\nu_i)|^2$ dominates,  and the spectral entanglement becomes maximal.  Thus, as the pump bandwidth $\sigma$ is decreased, the degree of spectral entanglement quantified by the Schmidt number $K$ increases while the heralded single-photon purity $K^{-1}$ decreases.    We have verified this behavior experimentally by extracting the heralded single-photon purity  from the diagonal Fourier transform spectroscopy  measurement detailed above, for a number of different pump bandwidths.  The results of this measurement are shown in Fig.~\ref{fig:purity}, along with a theoretical curve, showing excellent agreement.

\subsection{Fibre Spectroscopy\label{sec:Fibre-Spectroscopy}}

\begin{figure}
\begin{centering}
\includegraphics[width=3.0in]{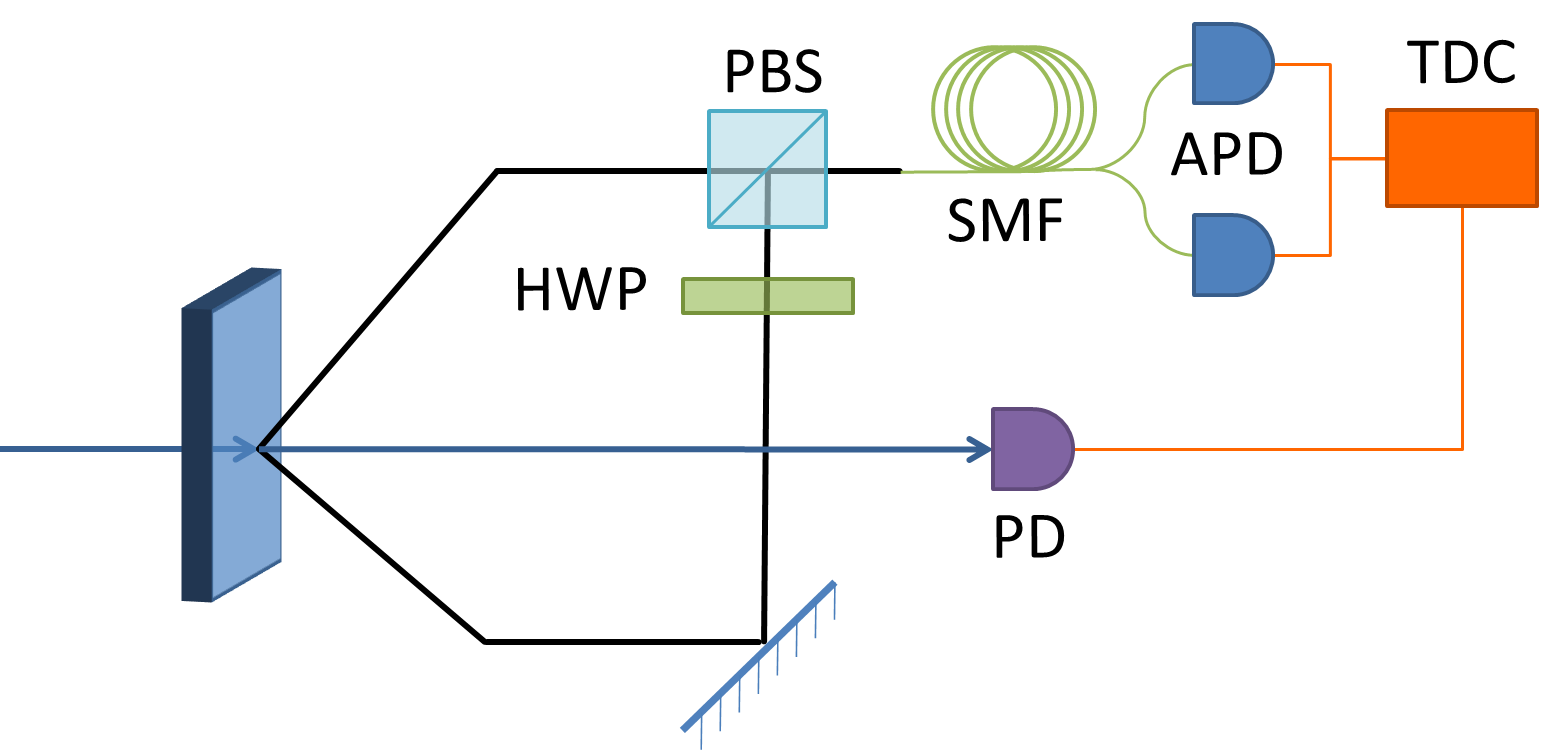}
\par\end{centering}
\caption[Fibre spectroscopy schematic]{\label{fig:780HPDispersion} Schematic diagram of fibre spectroscopy. A half-wave plate (HWP) and polarizing beamsplitter (PBS) are used to combine the signal and idler modes into a 400-m length of single-mode fibre (SMF). A fibre beamsplitter delivers light to two avalanche photodiodes (APD), which are analysed by a time-to-digital converter (TDC) together with a synchronization signal from the pump via a photodiode (PD).}
\end{figure}

As mentioned previously, interferometry is not the only method possible for transforming the spectral information of the two-photon state into a measurable form.  Another promising technique exploits dispersion in an optical fibre in order to map frequency components  into resolvable times of detection \cite{Avenhaus2009}. This exploits the property that short wavelengths travel more slowly than long wavelengths in an optical medium (with ordinary dispersion). The experimental schematic is shown in Fig. \ref{fig:780HPDispersion}. Our fibre (Nufern 780HP) has a dispersion of approximately -120 ps/nm/km for light near 810 nm. For example, in our $400$-m length of fibre, a photon with a wavelength of 809 nm will be delayed by approximately 50 ps compared to a photon at 810nm. Thus, measuring the time of arrival of a photon determines its wavelength, assuming that the relative delay exceeds the detector timing jitter. Higher resolution can be achieved using longer fibre lengths, but at the cost of greater loss; alternatively, media with higher dispersion can be used, as was shown recently through the use of chirped fibre Bragg gratings \cite{Davis2017}.

Detection is accomplished with two Micro Photon Devices avalanche photodiodes (APDs) with custom circuitry, chosen for low jitter (see Sec. \ref{sec:calibration}) \cite{stipcevic}. The transit time difference between the signal and idler photons in the single-mode fibre is on the order of 1ns, while the dead-time of each APD is over 20 ns; for each particular pair, there is a 50\% chance that the photons will arrive at different APDs and are thus counted in coincidence, and a 50\% chance that the photons will arrive at the same APD and are not counted in coincidence. This could be improved by using two separate lengths of fibre, but resources are typically better spent in obtaining one longer length of fibre, as the fibre length determines the amount of dispersion experienced by the photons, and thus the resolution of the measurement. The output pulses from the detectors are registered by an Agilent U1051A time-to-digital converter (`time-tagger'), with 50 ps time-bin resolution, and counted in coincidence. The coincidence signal is referenced to an Electro-Optics ET-2030 photodiode measuring the pump, also registered by the time-tagger.

The time-tagger is so named because it `tags' input pulses on a number of channels with a time of arrival. This allows convenient digital post-processing of events, with the ability to find coincidences between different channels with arbitrary intra-channel temporal delays and with arbitrary coincidence window widths; the specific configuration used may be freely chosen after the data are collected. The result is a very powerful tool for applications ranging from singles and coincidence counting, to correlation-based jitter measurements (Sec. \ref{sec:calibration}), to complex multi-fold coincidences (Sec. \ref{sec:fibre-measurement}).

\subsubsection{Fibre Spectroscopy Calibration\label{sec:calibration}}

In fibre spectroscopy, the timing jitter of the detection electronics limits the accuracy of the measurement. A high-jitter detector will lead to an uncertainty in the frequency-time relationship. This limits the resolution of the measurement, leading to a more circular joint spectrum and an artificially lower measured Schmidt number. Thus, calibration is essential to determining the measurement accuracy; we perform this by first characterizing the system jitter in the context of a spectroscopic measurement of a calibrated classical source.

This calibration includes using the time-tagger to measure the distribution of response times of each component in the experimental setup. We measured APDs from Perkin-Elmer, ID Quantique, and Micro Photon Devices, which had FWHM jitters of 358 ps, 70 ps, and 174 ps, respectively. The best-performing detectors were MPD avalanche photodiodes (with custom circuitry from Mario Stipcevic), which have a reasonably low jitter of 174 ps and do not have a significant non-Gaussian tail. Although the ID Quantique detectors have a lower FWHM jitter, they also have a long tail with a significant portion of the total power (over 50\% outside of the FWHM as compared to approximately 24\% for a Gaussian response), which severely reduces the resolution of the measurement. The time response of the time-tagger itself was measured using the correlation method, with a function generator simultaneously triggering two channels on the time-tagger. This yielded a FWHM jitter of 64 ps, which is deconvolved from all other jitter measurements. Additionally, the time response of the Electro-Optics photodiode was measured, yielding a FWHM jitter of 203 ps. Adding the jitter values in quadrature gives a total-system jitter of $\sqrt{203^2 + 174^2 + 64^2} = 275$ ps with the MPD detectors.

\begin{figure}
\begin{centering}
\includegraphics[width=3.0in]{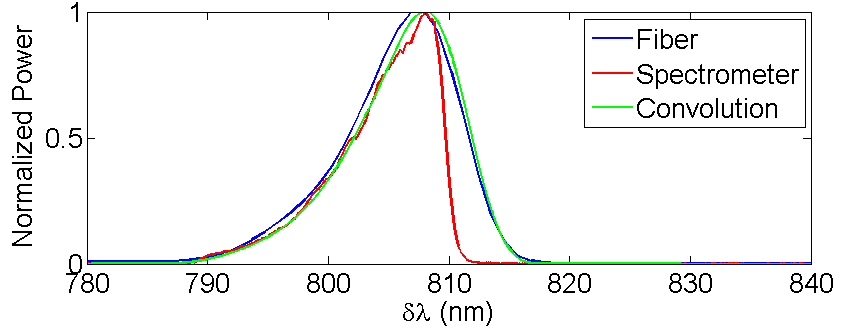}
\par\end{centering}
\caption[Fibre dispersion calibration]{\label{fig:fibre-calibration}Normalized spectral power density measurements under 810 nm low-pass filtering. The measured spectrum from an Ocean Optics HR2000 spectrometer is shown in red. The inferred spectrum from the fibre spectroscopy is shown in blue, re-centred on the peak of the spectrometer measurement (as the measurement only gives wavelength relative to an arbitrary reference). The green line shows the convolution of the spectrometer data with a 275 ps FWHM Gaussian, i.e., the measured timing jitter of our system. }
\end{figure}

We can compare this jitter with measurements obtained using fibre spectroscopy on a known calibration source. This measurement is similar to that pictured in Fig. \ref{fig:780HPDispersion} except that instead of an SPDC source and two APDs measuring in coincidence, there is only the pump, which is sent to a photodiode and a single APD. Additionally, the pump is optionally filtered by a low-pass filter to observe the response to a sharp spectral cutoff. The time response is determined by looking at the correlation between the APD and the photodiode, with the photodiode serving as a fixed point of reference. This is converted to a spectrum by multiplying by a conversion factor determined by the length and dispersion of the fibre, and a manual offset which varies depending on delays present in the experiment, but which does not depend on the shape of the spectrum. Finally, the inferred spectrum is compared to that measured directly with an Ocean Optics HR2000 spectrometer, which is also mathematically convolved with a Gaussian equivalent to a system jitter of 275 ps. Fig. \ref{fig:fibre-calibration} shows the results of this measurement, which exhibits excellent agreement between the directly measured spectrum and the fibre-spectroscopy-measured spectrum, given the characterized jitter of the system.

\subsubsection{Fibre Spectroscopy Measurement and Simulation for the SPDC Source\label{sec:fibre-measurement}}

\begin{figure}
\begin{centering}
(a)\includegraphics[width=1.4in]{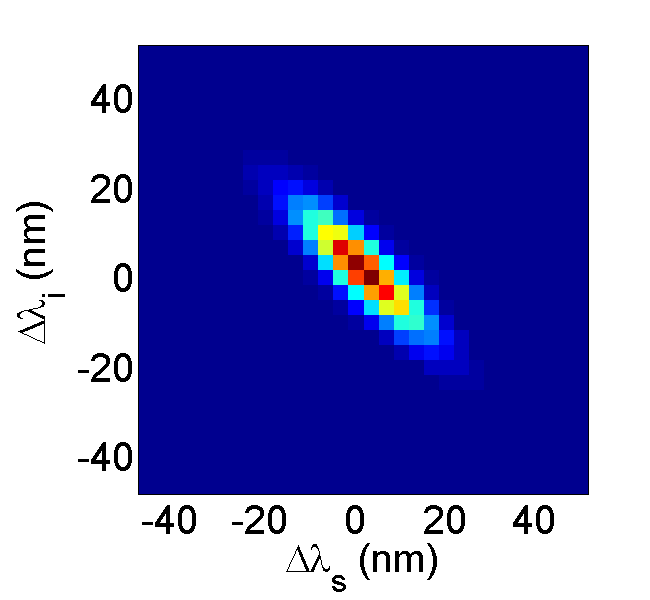}(b)\includegraphics[width=1.4in]{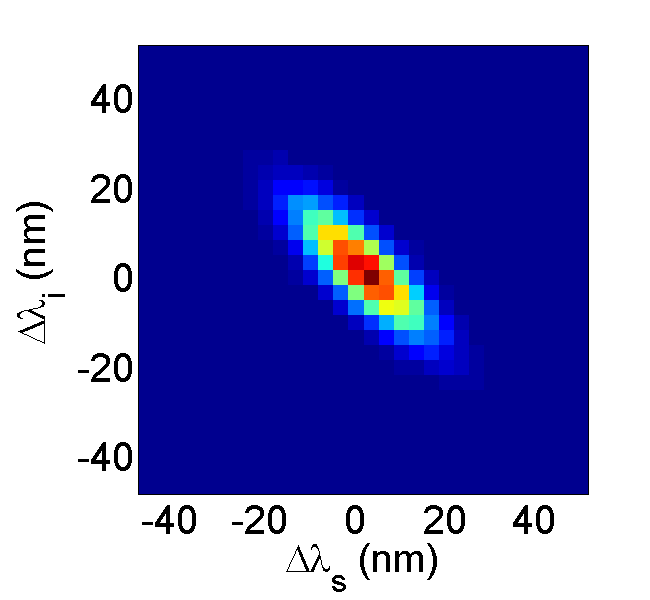}\\
(c)\includegraphics[width=1.4in]{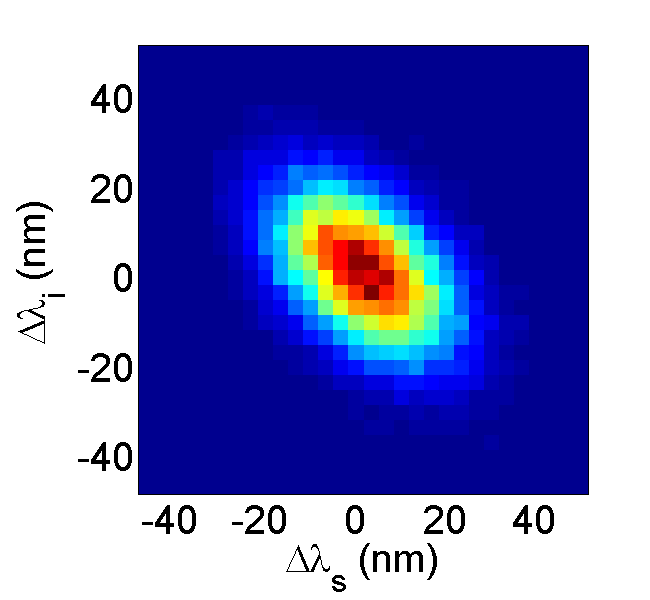}
\par\end{centering}
\caption[Simulation of fibre spectroscopy jitter]{\label{fig:simulated-fibre}Monte Carlo simulations of fibre spectroscopy with a detection timing jitter of (a) 50 ps, (b) 275 ps, and (c) 1 ns. Simulated detector photons are detected with a Gaussian random jitter and binned into a 2D histogram with 50 ps time bin widths (the bin width of our Agilent U1051A Time-to-Digital Converter). The true Schmidt number of this simulated source is 2.0, and the Schmidt numbers after the application of jitter are (a) $1.98 \pm 0.02$, (b) $1.68 \pm 0.01$, (c) $1.14 \pm 0.003$. A Richardson-Lucy deconvolution algorithm can be applied to attempt to remove the effect of the jitter, but this compensation breaks down for large jitters, resulting in (a) $1.98 \pm 0.02$ (b) $1.99 \pm 0.02$, (c) $1.20 \pm 0.007$. The simulation was performed with 32,000 simulated photon pairs, a fairly typical number for our experimental measurements.}
\end{figure}

The effect of jitter can be simulated using a Monte Carlo technique, with photons drawn from an assumed spectrum receiving random jitters to produce a transformed spectrum, as shown in Fig. \ref{fig:simulated-fibre}.\footnote{The system jitter is slightly more nuanced for a two-dimensional measurement; the photodiode jitter is correlated with both the signal and idler detection, as both are referenced to the photodiode signal. This shared reference leads to the signal and idler detection times appearing more correlated than they actually are. This manifests in the positive diagonal direction in Fig. \ref{fig:simulated-fibre}, which corresponds to the \emph{sum} of the two detection times, and therefore the jitter of the reference does not cancel. This contrasts with the anti-diagonal direction which corresponds to the difference of detection times, where the photodiode jitter does cancel. However, Monte Carlo simulations of this effect indicate that it is minor, leading to a purity bias of no more than 0.01 in our measurements.} This allows us to estimate the uncertainty of the experimental measurements. We also attempted to remove the detector response using a Richardson-Lucy \footnote{Richardson-Lucy is an iterative maximum-likelihood-based deconvolution algorithm developed for image processing, specifically for removing a Gaussian blur from an image. Initially, we attempted to use a naive matrix-inversion-based deconvolution, but this proved to be numerically unstable for our application due to the high condition number of the joint-spectral matrix.} deconvolution algorithm \cite{RICHARDSON1972}. This is reasonably successful for removing a relatively small jitter, but breaks down as jitter increases. As a rough rule of thumb, deconvolution is unnecessary if the jitter is less than half as large as the smallest spectral feature (i.e., the narrow axis of the ellipses of Fig. \ref{fig:simulated-fibre}), and is unreliable if the jitter is more than 2-4 times larger than the smallest spectral feature (the latter criterion depends on the amount of noise and the shape of the features). Jitter on the order of the feature size is a `sweet spot' for applying deconvolution.

\begin{figure}
\centering
(a)\includegraphics[width=1.357in]{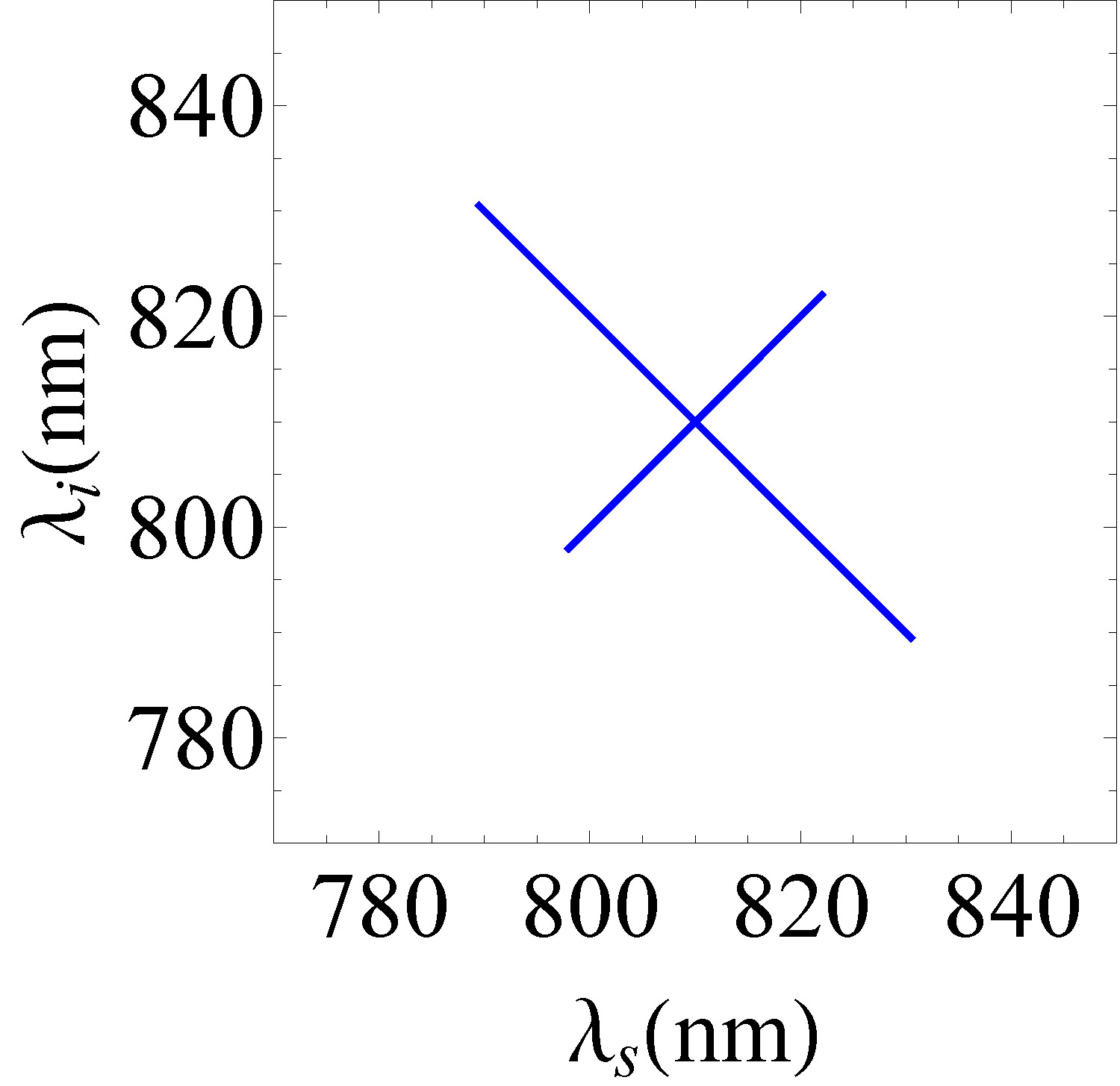}
(b)\includegraphics[width=1.423in]{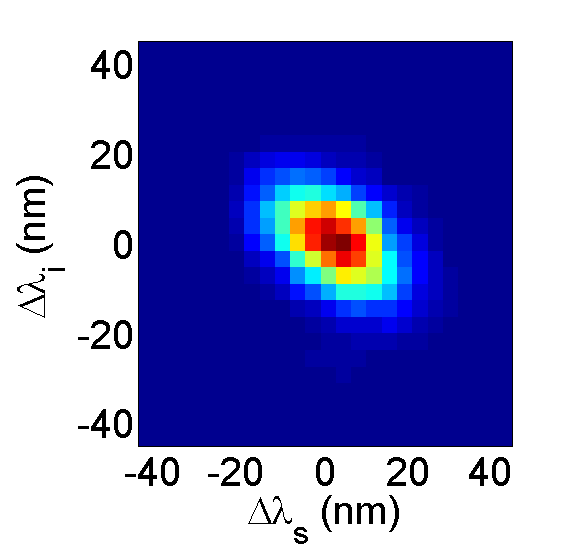}
(c)\includegraphics[width=1.478in]{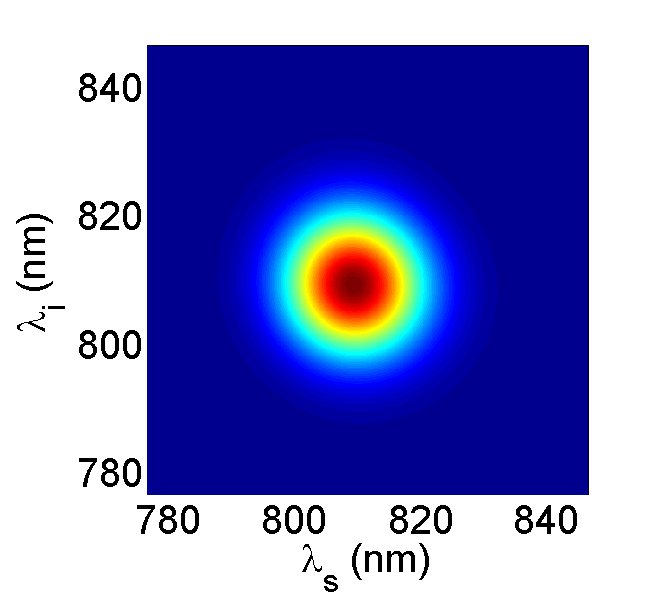}
\caption[Comparison of joint spectrum measurements]{\label{fig:JSI-comparison}Comparison of measurements on a source with no spectral filtering and an 8 nm bandwidth pump using (a) diagonal Fourier spectroscopy (b) fibre spectroscopy and (c) a theoretical simulation showing ideal behavior. Corresponding purities are (a) $0.88\pm0.02$, (b) $0.87\pm0.03$, and (c) 0.998. }
\end{figure}

To perform the full joint spectral measurement, we need to perform a complex multi-channel correlation. This is not a typical use of the time-tagger, and is not directly supported by the default configuration of the driver software, so some modifications must be made. The basic idea of the measurement is to count in three-fold coincidence between the APDs and the photodiode, with each triple-coincidence event being registered as an event with $\Delta\lambda_s \approx c(t_s - t_{pd})$ and $\Delta\lambda_i \approx c(t_i - t_{pd})$, where $c$ is the speed of light and $t_s$, $t_i$, and $t_{pd}$ are the timestamps of the signal photon, idler photon, and photodiode pulse, respectively. The set of all such events can be binned into a two-dimensional relative-timing histogram, then converted to wavelength to produce a joint spectrum similar to those of Fig.~\ref{fig:simulated-fibre} (see Fig. \ref{fig:JSI-comparison}). Finally, the Schmidt decomposition can be applied to this histogram to determine the Schmidt number as in Sec. \ref{sub:schmidt-modes}.

Unfortunately, directly accessing these times of arrival on all three channels is not practical given the extremely large number of time-tags involved. Instead, we make use of a fast and efficient multi-coincidence routine in custom driver software. This allows the fast construction of the desired histogram by the following method: first, we observe that each bin in the histogram corresponds to a three-fold coincidence with offsets $\Delta t_s$ and $\Delta t_i$ and window size equal to the histogram bin width. Then, by iterating over all offsets in the desired range, we can construct the desired histogram. A comparison of the results of this technique with Fourier spectroscopy measurements  and our simulations is shown in Fig. \ref{fig:JSI-comparison}. The two measurements agree closely with each other, but exhibit more correlation than the ideal simulated source. We believe this discrepancy arises from experimental factors not included in simulation, particularly temporal and spatial chirp in the pump.

\subsection{Stimulated-Emission-Based Measurement Technique\label{sec:seb}}

\begin{figure}
\begin{center}
\includegraphics[width=3in]{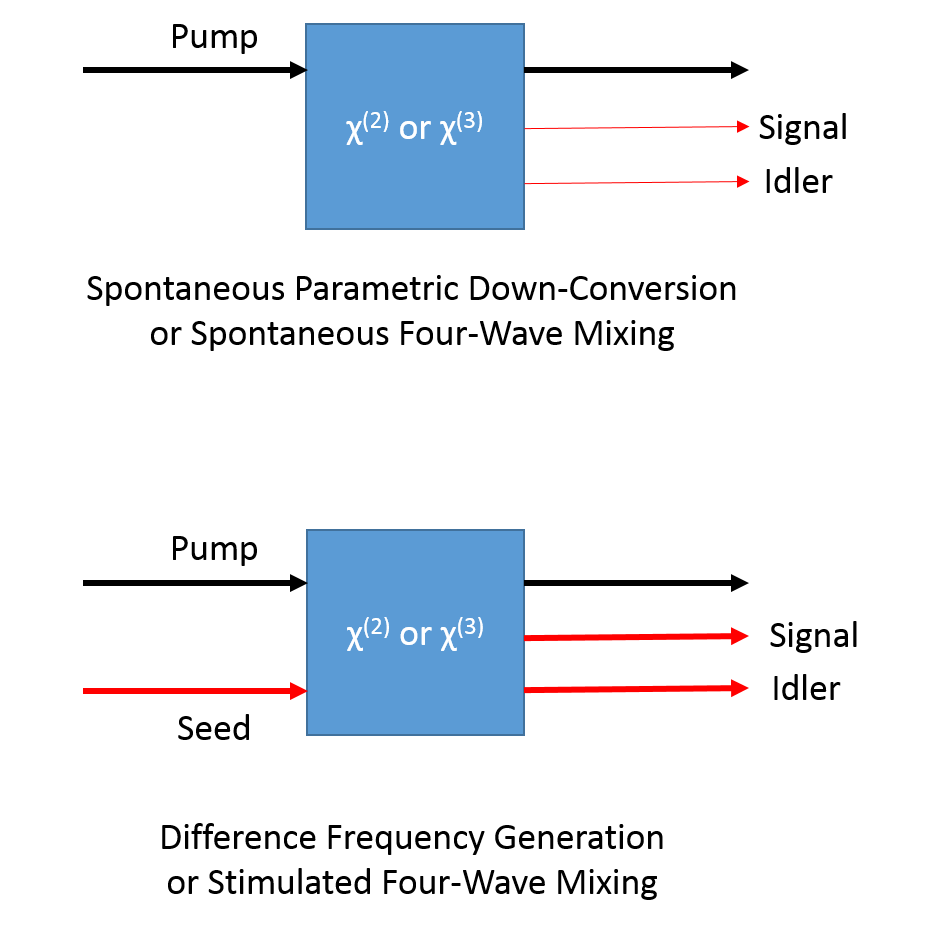}
\end{center}
\caption[Stimulated-emission-based measurement]{\label{fig:StimulatedAnalogues}Schematic of spontaneous processes and their stimulated analogues. Line thickness represents intensity.}
\end{figure}

Stimulated-emission-based measurement of the JSI relies on the relationship between the spontaneous process and its corresponding stimulated process \cite{Liscidini2013}. Specifically, difference-frequency generation can be used to characterize an SPDC source \cite{Eckstein2014}, and stimulated four-wave mixing can be used to characterize an SFWM source \cite{Fang2014}; the relation between the stimulated and spontaneous processes is depicted in Fig.~\ref{fig:StimulatedAnalogues}.  We can see that the spontaneous and stimulated spectra are related by comparing the expected number of photons from each process \cite{Liscidini2013}:
\begin{equation}
\frac{\braket{n_{\sigma_{s}\mathbf{k}_{s}}}_{A_{\sigma_{i}\mathbf{k}_{i}}}}{\braket{n_{\sigma_{s}\mathbf{k}_{s}}n_{\sigma_{i}\mathbf{k}_{i}}}} \approx \left |{A_{\sigma_{i}\mathbf{k}_{i}}}\right |^2
\label{eq:linear}
\end{equation}
where $\braket{n_{\sigma_{s}\mathbf{k}_{s}}}_{A_{\sigma_{i}\mathbf{k}_{i}}}$ is the average number of signal photons with polarization $\sigma_{s}$ and wave vector $\mathbf{k}_{s}$ stimulated by an idler seed with polarization $\sigma_{i}$ and wave vector $\mathbf{k}_{i}$, $\braket{n_{\sigma_{s}\mathbf{k}_{s}}n_{\sigma_{i}\mathbf{k}_{i}}}$ is the average number of spontaneously generated photon pairs, and $\left |{A_{\sigma_{i}\mathbf{k}_{i}}}\right |^2$ is the average photon number of the coherent seed pulse. Thus the number of photons resulting from a spontaneous process is directly proportional to the number of photons detected in its corresponding stimulated process, with the proportionality factor being the number of photons in the stimulating laser. If we use a mW-power continuous wave (CW) stimulating laser, this number is many orders of magnitude higher than $n_{\sigma, \sigma'}$ alone, leading to a significantly shorter collection time and a significantly higher signal-to-noise ratio than techniques based on coincidence counting. Note that in experiment the dependence of the stimulated signal power on seed power should be confirmed to be in the linear regime for the seed powers used \cite{Fang2014}. Note also that the stimulated emission technique is not sensitive to other processes that may generate noise background, such as Raman scattering; thus, such contributions should be characterized separately.

\subsubsection{Stimulated-Emission-Based Measurement for SFWM Sources}

\begin{figure}
\begin{center}
\includegraphics[width=3.4in]{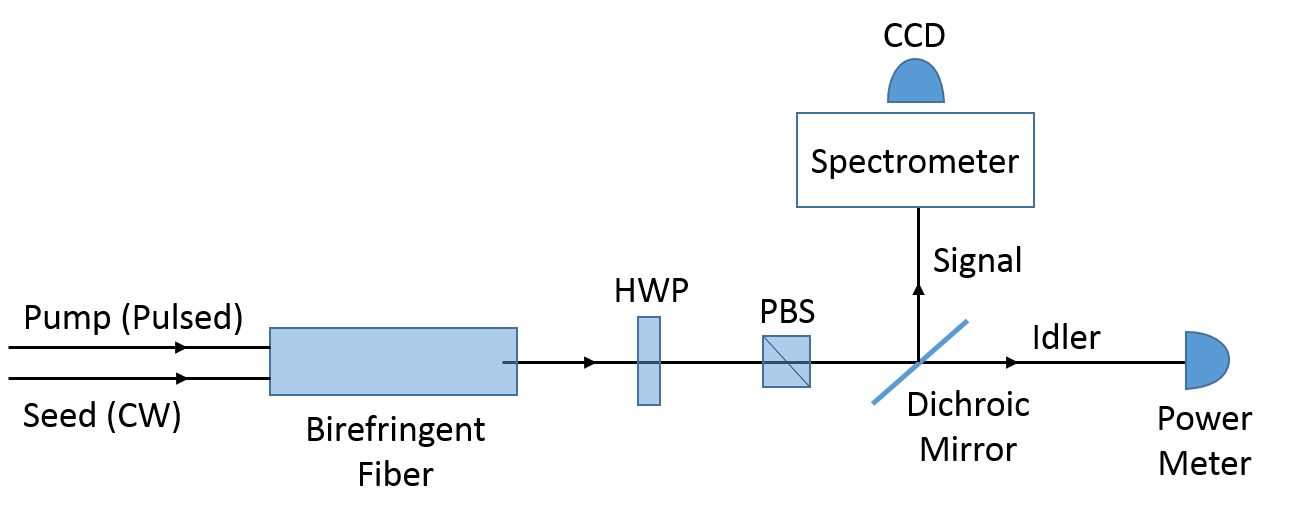}
\end{center}
\caption[Stimulated-emission-based measurement setup]{\label{fig:stimexpt}Experimental setup for the stimulated-emission-based measurement of the JSI of an optical fibre SFWM source. A half-wave plate (HWP) and polarizing beamsplitter (PBS) are used to filter out the pump.}
\end{figure}

The experimental setup for performing the stimulated-emission-based measurement on SFWM sources is shown in Fig.~\ref{fig:stimexpt}. A CW laser (the `seed') is coupled into the fibre in addition to the pulsed pump. The seed laser has a bandwidth of $\sim 30$GHz and its centre frequency is incrementally scanned over the frequency range of idler photons produced in the spontaneous process so a stimulated four-wave mixing process is driven and a stimulated signal beam is produced. The stimulated signal is sent into a standard spectrometer with a CCD that records the signal's spectrum at each seed frequency. A complete scan for this source takes approximately 15 minutes. The results for bow-tie polarization-maintaining fibre are shown in Fig.~\ref{fig:fibremeas}(b). When compared to the monochromator result shown in Fig.~\ref{fig:fibremeas}(a), the signal-to-noise ratio is clearly much higher and the resolution is much better. The purity calculated from the joint spectral intensity measured using stimulated emission is $0.8322\pm0.0004$.

\begin{figure}
\begin{center}
\includegraphics[width=3.3in]{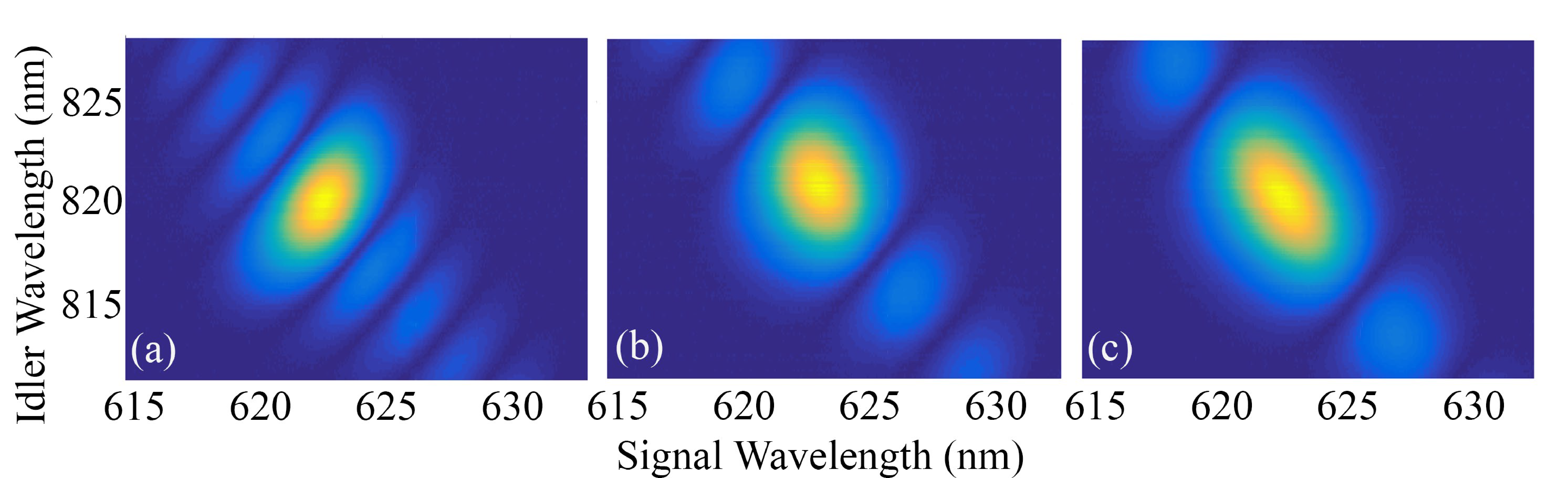}
\end{center}
\caption[SFWM Engineering]{\label{fig:sebeng}Joint spectral intensities from stimulated-emission-based measurements of panda-fibre SFWM sources of three different lengths: (a) 2.6 cm, (b) 1.6 cm, and (c) 1.1 cm.}
\end{figure}

As the stimulated emission technique is relatively swift, it allows the efficient testing of engineered sources. To demonstrate this we show in Fig.~\ref{fig:sebeng} the JSI of three SFWM sources in panda-type polarization-maintaining fibre for three different fibre lengths that exhibit correlated (Fig.~\ref{fig:sebeng}(a)), almost uncorrelated (Fig.~\ref{fig:sebeng}(b)), and anti-correlated (Fig.~\ref{fig:sebeng}(b)) JSI for fibres of length 2.6 cm, 1.6 cm, and 1.1 cm, respectively. The purities calculated from the measured JSI of these sources assuming a flat joint spectral phase are $0.8018\pm0.0003$, $0.8975\pm0.0001$, and $0.8529\pm0.0002$, respectively.

\section{Heralded Single-Photon Purity Measurement\label{chap:measuring-purity}}

In this section we measure the heralded single-photon purity, which is phase-dependent and which in many situations is the quantity of interest, in two ways: indirectly, by taking advantage of the statistical properties of SPDC and SFWM (Sec. \ref{sec:correlation-function}), and directly by interfering heralded single photons from two SPDC sources (Sec. \ref{sec:two-source-hom}).

\subsection{Correlation Function Measurement\label{sec:correlation-function}}

The second-order correlation function $g^{(2)}(t_1,t_2)$ represents the joint probability of detecting one photon at a time $t_1$ and another photon at a time $t_2$. As we will discuss in this section, the time-integrated $g^{(2)}$ function can be used to determine the purity of the heralded single-photon quantum state. The second-order correlation function $g^{(2)}(t_1,t_2)$ may be expressed, in terms of time-dependent photon number operators $\hat{n}_\mu(t)=\hat{a}_\mu^\dagger(t)\hat{a}_\mu(t)$, with $\mu=1,2$, as

\begin{equation}
g^{(2)}(t_1,t_2) = \frac{\left\langle : \hat{n}_1(t_1)\hat{n}_2(t_2): \right\rangle}{\left\langle \hat{n}_1(t_1)\right\rangle\left\langle \hat{n}_2(t_2)\right\rangle},\label{eq:g2-n}
\end{equation}

\noindent where $\langle ... \rangle$ indicates normal ordering.

\begin{figure}
\begin{flushleft}
\centering
(a)\includegraphics[width=2in]{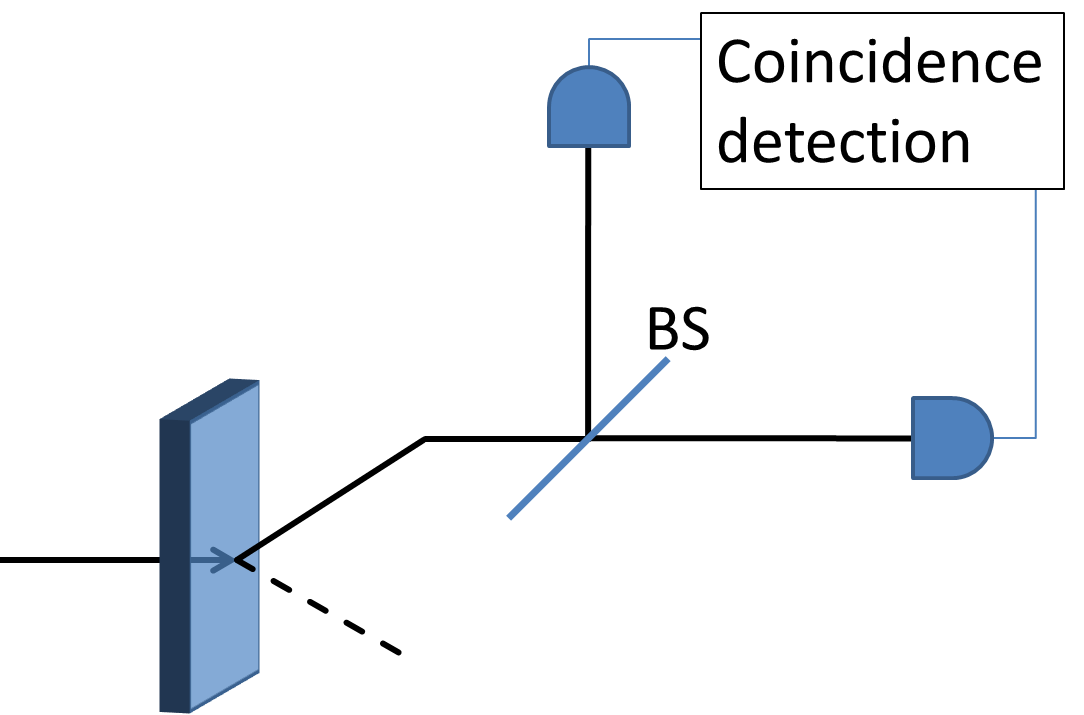}
(b)\includegraphics[width=1.6in]{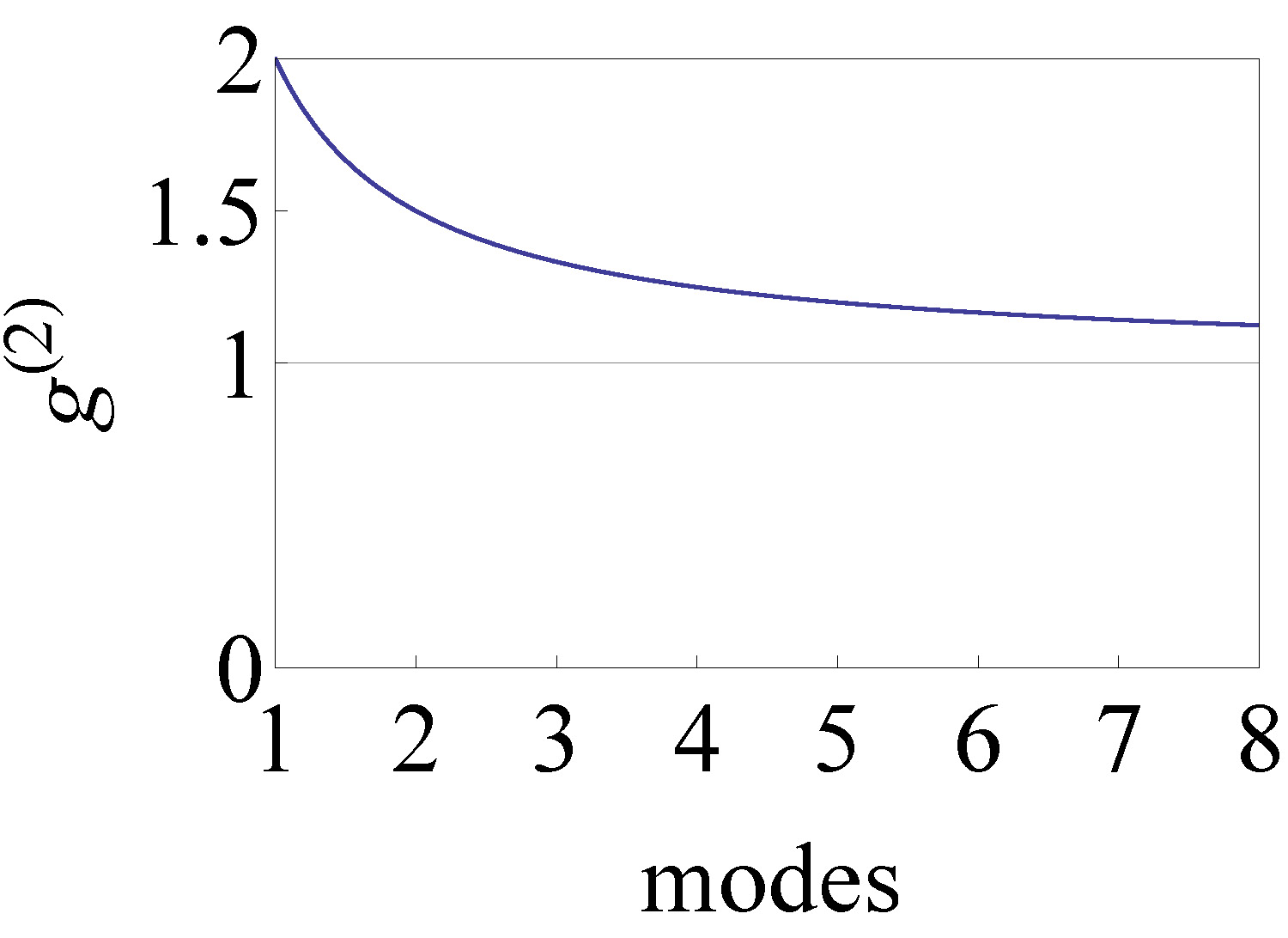}
\par\end{flushleft}
\caption[Correlation function measurement scheme]{\label{fig:g2-scheme}(a) Setup for measuring $g^{(2)}$ in one arm of SPDC using a Hanbury Brown-Twiss interferometer and (b) calculated relationship between $g^{(2)}$ for a source with $K$ (effective) thermal modes.}
\end{figure}

In our case we are interested in measuring the $g^{(2)}$ function of the signal or idler mode of an SPDC source.  This measurement can be accomplished by sending the desired mode to a Hanbury Brown-Twiss interferometer \cite{Brown1956}, as shown schematically in Fig.~\ref{fig:g2-scheme}(a), where the $1$ and $2$ labels correspond to the output ports of the beamsplitter, each leading to an avalanche photodiode.   From Eq. \ref{eq:g2-n}, the $g^{(2)}(t_1,t_2)$ function is given as the ratio of the time-resolved coincidence rate between the two detectors divided by the product of the time-resolved single-channel detection rates in each of the two detectors.

The possibility of temporally resolving the $g^{(2)}$ function would hinge on a fast detector response, as compared to the coherence time.  The latter condition is  \emph{not} fulfilled in our experimental apparatus: our broadband pulses have coherence times in the tens of femtoseconds, while the APD's have typical response times in the hundreds of picoseconds!   Thus, in the experimentally realistic case of a slow detector response, we instead measure the time-integrated correlation function $g^{(2)}$\cite{Christ2011}, expressed as


\begin{equation}
g^{(2)} = \frac{\int d t_1 \int d t_2\left\langle : \hat{n}_1(t_1)\hat{n}_2(t_2): \right\rangle}{\int d t_1\left\langle \hat{n}_1(t_1)\right\rangle\int d t_2\left\langle \hat{n}_2(t_2)\right\rangle},
\end{equation}

\noindent where the integral is taken over the detection window, assumed to be long compared to the pulse duration. The time-integrated $g^{(2)}$ is sensitive to the distribution of Schmidt modes of the measured photons (although it does not provide detailed temporal information as $g^{(2)}(t_1,t_2)$ does) — this  leads to its utility for determining the purity of the heralded single-photon state, as we discuss in further detail below.

In carrying out this measurement with single-photon detectors,  one must either take care that the probability of multiple photons arriving during each detector's `dead-time' is negligible, or alternatively employ photon-number-resolving detectors. We satisfy the first condition by using a signal/idler average photon number per pump pulse which is much less than 1 (approximately 0.001).

Recall that when carrying out a Schmidt decomposition, the two-photon state may be written in terms of the Schmidt eigenvalues $\lambda_j$ and the Schmidt annihilation operators $\hat{A}_j$ for the signal mode and $\hat{B}_j$ for the idler mode, as

\begin{equation}
|\Psi\rangle=\sum\limits_{j=1}^\infty \sqrt{\lambda_j} A_j^\dagger B_j^\dagger |0\rangle.
\end{equation}

In Ref. \cite{Christ2011} it was shown that this time-integrated $g^{(2)}$ function may be written, for the signal mode with $\hat{N}_s\equiv\hat{A}_{i}^\dagger \hat{A}_{i}$,  as

\begin{equation}
g^{(2)}=\frac{\left\langle :\left(\sum\limits_{i=1}^{\infty} \hat{N}_s \right)^{2}:\right\rangle }{\left\langle \sum\limits_{i=1}^{\infty} \hat{N}_s\right\rangle ^{2}}, \label{eq:g2-N}
\end{equation}

\noindent  and  an equivalent expression may be written with $\hat{N}_i\equiv\hat{B}_{i}^\dagger \hat{B}_{i}$ for the idler mode.  Note that for single-mode SPDC, i.e., for which each of the signal and idler waves occupies a single polarization/spatial/spectral mode, there is a single term in each of the two sums in Eq. \ref{eq:g2-N}, and each wave then has thermal statistics with $g^{(2)}=2$.    It is interesting that for an SPDC source which departs from being single-mode, the statistics for each of the signal and idler waves is no longer thermal.     Consider a concrete example of two independent polarization modes, H and V, each in a thermal state

\begin{multline}
g^{(2)}=\frac{\left\langle :\left(\hat{n}_{H}+\hat{n}_{V}\right)^{2}:\right\rangle }{\left\langle \hat{n}_{H}+\hat{n}_{V}\right\rangle ^{2}}
=\frac{\left\langle :\hat{n}_{H}^{2}+2\hat{n}_{H}\hat{n}_{V}+\hat{n}_{V}^{2}:\right\rangle }{\left\langle \hat{n}_{H}\right\rangle ^{2}+2\left\langle \hat{n}_{H}\right\rangle \left\langle \hat{n}_{V}\right\rangle +\left\langle \hat{n}_{V}\right\rangle ^{2}}\\
=\frac{\left\langle :\hat{n}_{H}^{2}:\right\rangle +2\left\langle :\hat{n}_{H}\hat{n}_{V}:\right\rangle +\left\langle: \hat{n}_{V}^{2}:\right\rangle }{4\left\langle \hat{n}\right\rangle ^{2}}\\
=\frac{1}{2}\left(\frac{\left\langle: \hat{n}^{2} :\right\rangle }{\left\langle \hat{n}\right\rangle ^{2}}+\frac{\left\langle :\hat{n}_{H}\hat{n}_{V}:\right\rangle }{\left\langle \hat{n}\right\rangle ^{2}}\right),
\end{multline}

\noindent where we have assumed for simplicity that the two modes are equally occupied, so that $\left\langle\hat{n}_{H}\right\rangle^2 = \left\langle\hat{n}_{V}\right\rangle^2 \equiv \left\langle\hat{n}\right\rangle^2$  and $\left\langle:\hat{n}_{H}^2:\right\rangle = \left\langle:\hat{n}_{V}^2:\right\rangle \equiv \left\langle:\hat{n}^2:\right\rangle$. The first term represents a single-mode thermal state, and has a value of 2. The second term, however, depends on correlations between the H and V modes. These are independent, uncorrelated modes, and so no photon bunching occurs, and this term takes a value of 1. Thus, the sum for our two-mode state is $g^{(2)}=3/2$.

This generalizes, for an SPDC source with a Schmidt number $K$, which corresponds to a source with $K$ effective thermal modes, to the following result
\begin{equation}
g^{(2)} = 1 + \frac{1}{K},
\end{equation}

\noindent which is shown in Fig.~\ref{fig:g2-scheme}(b). Therefore, measuring $g^{(2)}$ directly determines the effective number of modes as $K = 1/(g^{(2)}-1) $, and the heralded single-photon purity as $P=1/K=g^{(2)}-1$. It is remarkable that information about the degree of entanglement in the photon pair ($K$) and about the single-photon purity, can be obtained from measuring only one arm of an SPDC source or an SFWM source. \footnote{For this to be exactly true, the complementary modes must be collected in the other arm. Otherwise, it is the (smaller) set of modes collected by both arms that determines heralded single-photon purity. This effect can cause the single-arm measurement to underestimate the coincidence post-selected purity.} Additionally, unlike the previously described methods, this measurement is sensitive to the total number of independent modes in the Schmidt decomposition, and in particular it is sensitive to the joint spectral phase (e.g., which could result from chirp in the pump).

\begin{figure}
\begin{centering}
\includegraphics[width=3.2in]{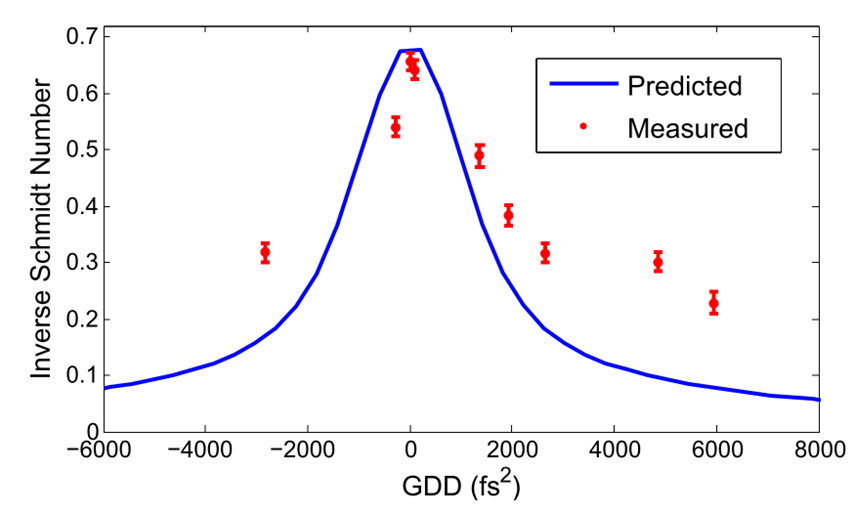}
\par\end{centering}
\caption[Correlation function measurement results]{\label{fig:g2-results}Values of inverse Schmidt number $1/K$, measured from the correlation function. This is equal to heralded single-photon purity or  $g^{(2)}(0)$ - 1. The horizontal axis shows a variable amount of group delay dispersion (GDD) applied to our pump using a prism pair compressor, controlling pump temporal chirp. A simple prediction based only on second- and third-order dispersion, where the third-order chirp was fixed to the value which optimizes agreement with the experimental results,  is shown in blue.}
\end{figure}

\subsubsection{Correlation Function Measurement of the SPDC source}

The features outlined in the previous section enable us to measure correlations present in the joint spectral phase; although we knew such correlations could be caused by a temporally chirped pump, we had no convenient way to measure this chirp, as our pump wavelength is outside the range accommodated by typical autocorrelation techniques. The $g^{(2)}$ measurement is sensitive to these phase correlations, and thus also indirectly describes the degree of temporal chirp in the pump. Fig.~\ref{fig:g2-results} shows the results of applying this technique to optimize dispersion compensation in the pump. With optimal dispersion compensation, we measure $g^{(2)} = 1.66 \pm 0.02$ without spectral filtering, and $g^{(2)} = 2.02 \pm 0.04$ with a 20 nm bandwidth filter; these values corresponds to a heralded single-photon purity of $0.66 \pm 0.02$ without spectral filtering and $1.02 \pm 0.04$ with spectral filtering.   The exact nature of the reduction in the purity in the absence of filtering, as compared with the values obtained with diagonal Fourier spectroscopy and/or fibre spectroscopy,  is not known, but it is likely due to a combination of higher-order dispersion in the pump and unintended filtering.  A model based on second- and third-order dispersion in the pump is shown in Fig.~\ref{fig:g2-results}; however, this model does not completely account for the effect observed.

\subsubsection{Correlation Function Measurement of the SFWM sources}\label{sec:CorrSFWM}

We performed correlation measurements on the bowtie-fibre SFWM source whose JSI is depicted in Fig.~\ref{fig:fibremeas}. The purity obtained from the $g^{(2)}$ value for this source is $0.63 \pm 0.02$, which is lower than the stimulated-emission-based measurement purity of 0.8325 and the monochromator purity of 0.80. We also measured $g^{(2)}$  values for the three panda-fibre SFWM sources whose JSI are depicted in Fig.~\ref{fig:sebeng}, to be $0.67 \pm 0.05$, $0.78\pm 0.07$, and $0.71\pm 0.07$, for 2.6cm, 1.6cm, and 1.1cm lengths of fibre, respectively. Our stimulated-emission-based measurements provide an upper bound for the purity because not all of the sidelobes in the JSI are measured, and because we did not extend the technique to measure the relative spectral phase between the signal and idler photons, which could contain correlations that degrade the purity. A summary of the results for the SFWM sources is presented in Table \ref{tab:Measurement-comparison-tableSFWM}.

\subsection{Two-Source Hong-Ou-Mandel Interference\label{sec:two-source-hom}}

\begin{figure}
\begin{centering}
\includegraphics[width=3.2in]{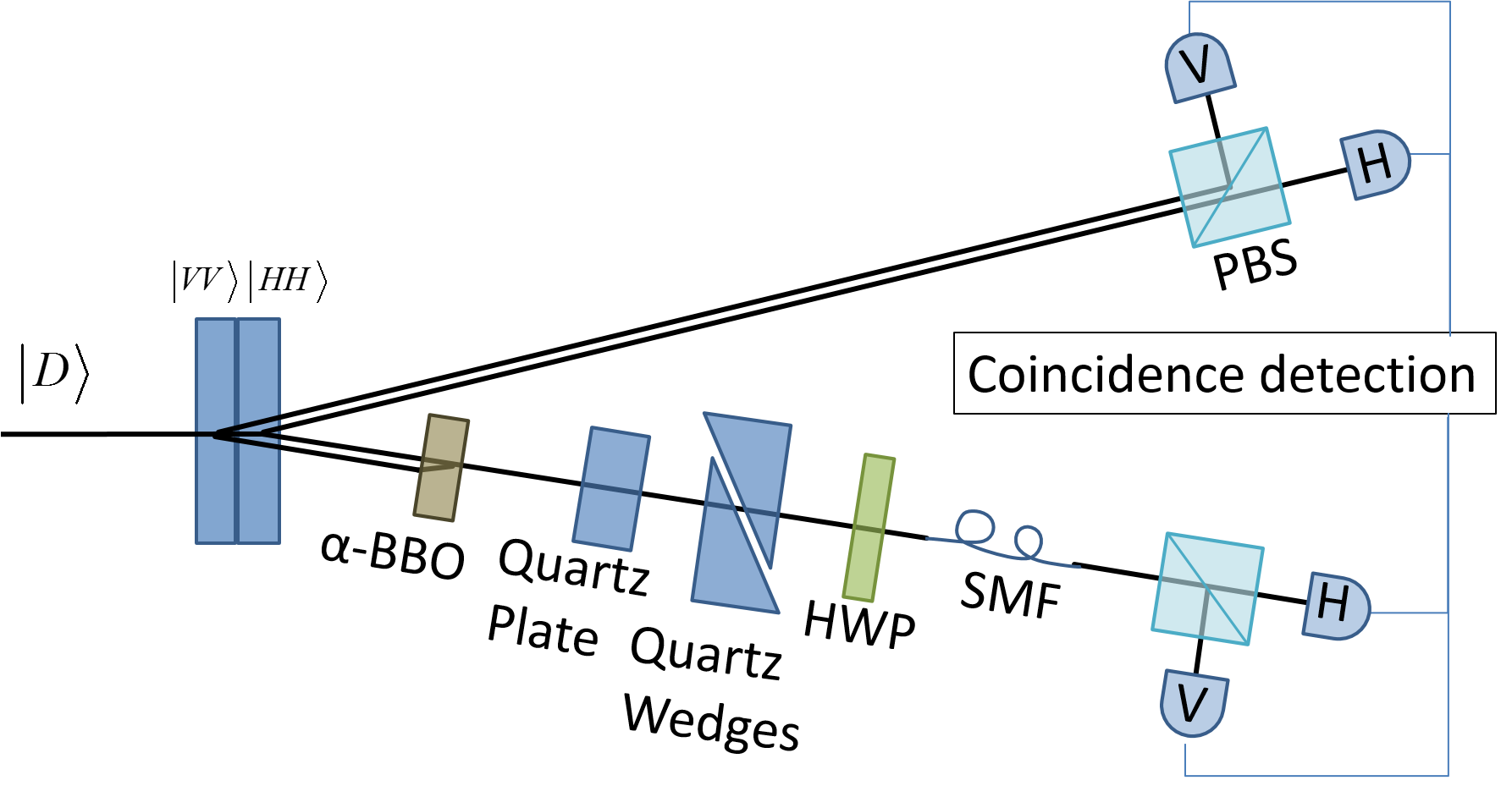}
\par\end{centering}
\caption[Two-source HOM interferometer]{\label{fig:HOMIsetup}Two-crystal common-path HOM interferometer. A diagonally polarized beam pumps two crystals orthogonally oriented to produce vertical and horizontal photon pairs. In the signal arm, these distinguishable photons are directed to two detectors, heralding the presence of one photon pair from each crystal. In the idler arm, we use transverse walk-off in birefringent $\alpha$-BBO to combine the photons into a single spatial mode. The quartz plate and quartz wedges compensate for the temporal delay between the photons, which are coupled into a single-mode fibre and rotated into the diagonal/anti-diagonal basis. An HOM dip can be observed as a suppression of four-fold coincidence counts as the temporal compensation is adjusted.}
\end{figure}

In many situations it is essential to interfere heralded single photons from a given source with other single photons from independent sources. The ability of single photons from distinct sources to interfere is governed by the single-photon purity; thus an interference experiment may be used for the determination of the purity.  This more direct route, in contrast to exploiting a spectral characterization of the source as in the cases of diagonal Fourier-transform and fibre spectroscopies described above, is considerably more time-consuming however, because it relies on the simultaneous emission (and detection) of two photon pairs, i.e., its intrinsically a four-photon experiment.

Recall that Hong-Ou-Mandel interference  relies on two single photons which impinge on a beamsplitter;  when the two scenarios which can lead to the two photons emerging from different output ports of the beamsplitter are indistinguishable, they interfere destructively and a null in the coincidence rate across these two output ports is expected. In a two-source HOM interferometer, the two intefering photons originate from independent sources; in our case the two sources are two type-I SPDC crystals with orthogonally oriented optic axes, as is commonly used for the generation of polarization-entangled photon pairs \cite{kwiat99}.   Our implementation of the two-source HOM interferometer employs a common-path polarization scheme with a similar motivation as in the case of polarization interferometer of Sec. \ref{sec:fourier-meas} \cite{Cohen2009}, i.e. in order to guarantee interferometric stability. Rather than being distinguished by their spatial mode, the two sources in fact share the same spatial mode, but have orthogonal polarization modes.

\begin{figure}
\begin{centering}
\includegraphics[width=3.2in]{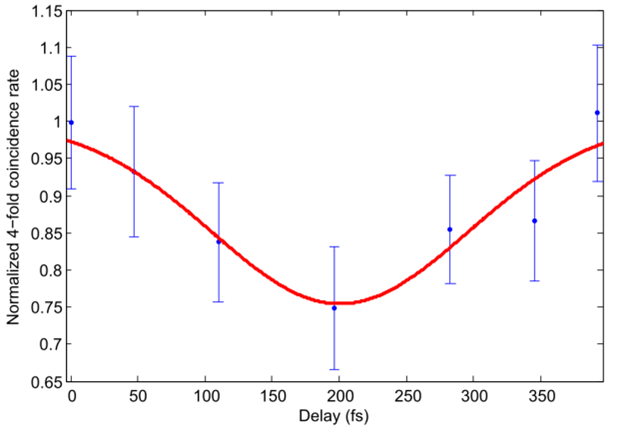}
(a)\\
\includegraphics[width=3.2in]{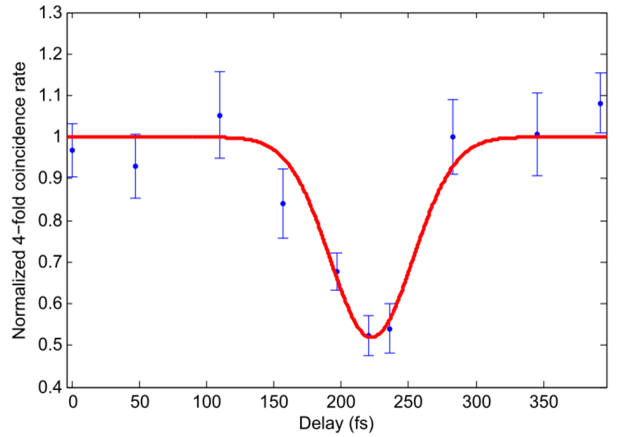}
(b)\\
\includegraphics[width=3.2in]{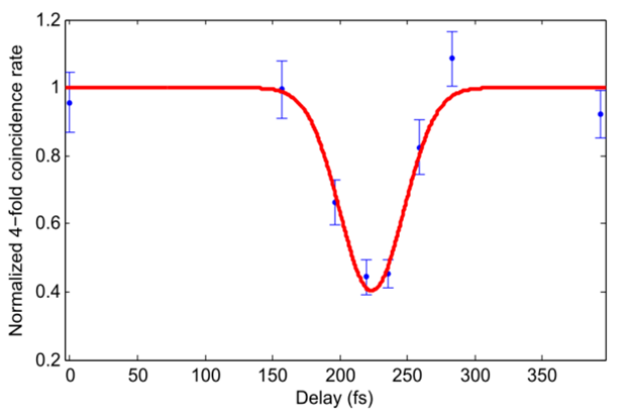}
(c)\\
\par\end{centering}
\caption[Results]{\label{fig:HOM-results} Results from two-source HOM with varying alignment and dispersion compensation. The visibilities of the HOM dips are
(a) $0.24\pm0.05$, (b) $0.48\pm 0.04$, and (c) $0.61\pm 0.05$. These show good agreement with
accompanying $g^{(2)}$ measurements, which imply a maximum visibility of (a) $0.26\pm 0.02$ and (b-c) $0.66\pm0.02$.  The low visibility in (a) is due to intentional misalignment of the dispersion compensation. The difference
between (b) and (c) is due to imperfect collection mode matching in (b), which was improved to obtain
the result in (c).   Red  lines shown indicate Gaussian fits to the data.}
\end{figure}

The experimental setup is shown in Fig.~\ref{fig:HOMIsetup}.  The two photons in the upper arm shown in the diagram are used as triggers; i.e., when a click is registered in both of the upper-arm detectors, a photon pair with orthogonal polarizations $| HV\rangle$ is heralded in the lower arm.  We observe HOM interference even though the two photons arrive on the same beamsplitter input port because we rotate the polarizations of the heralded photons into the diagonal (D)/ anti-diagonal (A) polarization basis and use a polarizing beamsplitter in the horizontal (H)/ vertical (V) basis.    The two photons exit on the same port of the polarizing beamsplitter, giving the characteristic HOM suppression of (four-fold) coincidence counts. The degree of indistinguishability of the heralded single photons, linked to  the HOM visibility $V$\footnote{Defined as $V \equiv 1-C_{min}$, where $C_{min}$ is the minimum count rate normalized by the baseline count rate.}, is related to heralded single-photon purity $P$ through the single-photon density operator $\hat{\rho}$ by

\begin{equation}
V=\mathrm{Tr} \left[ \hat{\rho}^2 \right] \equiv P.
\end{equation}

An important implementation detail is the use of birefringent walk-off in $\alpha$-BBO for spatial compensation in the idler arm.\footnote{One could instead implement spatial \emph{pre}-compensation on the pump, causing the two idler modes to directly overlap.} This removes which-crystal distinguishing information at the beamsplitter and allows two-source heralded single-photon interference.

Our measurement results under different conditions are shown in Fig.~\ref{fig:HOM-results}. In panel (a), we show a HOM dip with a relatively low visibility, resulting from intentional misalignment of the dispersion compensation, so that the pump carries a non-zero quadratic chirp.  In panel (b), we show a HOM dip resulting from imperfect collection mode matching, which was improved upon to obtain the result shown in panel (c).
In this last panel  we observe a maximum two-source HOM visibility of $0.61 \pm 0.05$, obtained without spectral filtering, which is consistent with the measured $g^{(2)}$-implied purity of $0.65 \pm 0.02$. Indeed, the $g^{(2)}$-implied purity is an upper limit on the HOM visibility, as the latter is sensitive not just to the heralded single photons being in a pure state, but also being in the \emph{same} pure state. That we are essentially able to reach this limit serves as an important cross-check between the information provided by these two measurements. It validates the use of the correlation function measurement as a proxy for heralded single-photon purity. This is useful because although the two-source HOM measurement provides unambiguous evidence for a lower bound on heralded single-photon purity, it is extremely time intensive. Each point on Fig.~\ref{fig:HOM-results} requires several hours of photon counting due to the low rate of two-pair events, leading to an entire day or more of data collection to obtain the full HOM dip. The correlation function measurement, on the other hand, can be performed with our source in under one hour.

\section{Conclusions}

\begin{table*}
\centering
\begin{tabular}{|c|c|c|c|c|}
\hline
                    & Fourier   & Dispersive  & Correlation & Two-Source \Tstrut\\
                    & Transform &  Fibre      & Function    & HOM \\    \hline%
Implied Purity
                    & 0.88$\pm$0.02
                    & 0.87$\pm$0.03
                    & 0.66$\pm$0.02
                    & 0.61$\pm$0.05 \Tstrut\\%
                    & (w/o~filters)
                    & (w/o~filters)
                    & (w/o~filters)
                    & (w/o~filters)\\%
                    & 0.99$\pm$0.01
                    & 0.995$\pm$0.04
                    & 1.02$\pm$0.02
                    & \\%
                    & (20nm~filters)
                    & (20nm~filters)
                    & (20nm~filters)
                    &\\ \hline             
\end{tabular}
\caption{Comparison of implied purity for each experimental characterization technique used for the SPDC source.\label{tab:Measurement-comparison-table}}
\end{table*}

\begin{table*}
\centering
\begin{tabular}{|m{1.5cm}|m{2cm}|m{4.5cm}|m{4cm}|}
\hline
& \multicolumn{1}{m{2cm}|}{\centering Scanning Monochromator} & \multicolumn{1}{m{4.5cm}|}{\centering Stimulated-Emission-Based} & \multicolumn{1}{m{4cm}|}{\centering Correlation Function} \\ \hline%
Implied      & 0.826$\pm$0.004     & 0.8322$\pm$0.0004~(bow-tie) 
                                                            & 0.63$\pm$0.02~(bow-tie) \\
Purity       &     (bow-tie)              & 0.8018$\pm$0.0003~(2.6\;cm) 
                                                            & 0.67$\pm$0.05~(2.6\;cm)\\
             &                   & 0.8975$\pm$0.0001~(1.6\;cm)
                                                            & 0.78$\pm$0.07~ (1.6\;cm)\\
             &                   & 0.8529$\pm$0.0002~(1.1\;cm)
                                                            & 0.71$\pm$0.07~(1.1\;cm)\\ \hline                                                                                                                        
\end{tabular}
\caption{Comparison of implied purity for each experimental characterization technique used for the SFWM sources. The fibre measured was panda-type unless otherwise noted.\label{tab:Measurement-comparison-tableSFWM}}
\end{table*}

\begin{table*}[h!]
\centering
\begin{tabular}{|p{20mm}|p{16mm}|p{15mm}|p{15mm}|p{15mm}|p{15mm}|p{15mm}|}
    \hline
    & \multicolumn{1}{L{15mm}|}{Scanning Monochromator} & \multicolumn{1}{L{15mm}|}{Diagonal Fourier Transform} & \multicolumn{1}{L{15mm}|}{Dispersive Fibre} & \multicolumn{1}{L{15mm}|}{Stimulated Emission} & \multicolumn{1}{L{15mm}|}{Correlation Function} & \multicolumn{1}{L{15mm}|}{Hong-Ou-Mandel} \Tstrut\\
    \hline
    Reconstruct JSI? & Yes & Yes & Yes & Yes & No & No \Tstrut\\
    \hline
    Phase-sensitive? & No & No & No & Yes* & Yes & Yes \Tstrut\\
    \hline
    Spectral Resolution & \multicolumn{1}{L{15mm}|}{0.20\;nm $\times$0.20\;nm} & N/A & \multicolumn{1}{L{15mm}|}{1\;nm $\times$1\;nm} & \multicolumn{1}{L{15mm}|}{0.06\;nm $\times$0.17\;nm} & N/A & N/A \Tstrut\\
    \hline
    Spectral resolution limited by & \multicolumn{1}{L{15mm}|}{spectrometer} & stage translation & electronic jitter & \multicolumn{1}{L{15mm}|}{spectrometer, seed scanning} & N/A & N/A\Tstrut\\
    \hline
    Peak Count Rate (counts/sec)&  0.57 & 2,236 & 2,058 & N/A & 0.1 & 0.1\Tstrut\\
    \hline
    Nominal Acquisition Time & 39 hrs & 700 s & 300 s & 200 s & 1 hr & 1 day\Tstrut\\
    \hline
    \multicolumn{1}{|L{20mm}|}{Acquisition time per bin} & 60 s & 1 s & 0.6 s & $10^{-3}$ s & N/A & N/A\Tstrut\\
    \hline
    Raw SNR & 6 & 47 & 45 & 198 & 20 & 9 \Tstrut\\
    \hline
    Scaled SNR & \multicolumn{1}{L{15mm}|}{$2.6$ s$^{-1}\mathrm{nm}^{-2}$} & 47 s$^{-1}$ & 75 s$^{-1}\mathrm{nm}^{-2}$ & \multicolumn{1}{L{15mm}|}{2$\times10^7$ s$^{-1}\mathrm{nm}^{-2}$} & \multicolumn{1}{L{15mm}|}{$6\times10^{-3}$ s$^{-1}$} & \multicolumn{1}{L{15mm}|}{$1\times10^{-4}$ s$^{-1}$}\Tstrut\\
    \hline
    \multicolumn{7}{l}{*Not shown in this work. See \cite{Jizan2016}.}\Tstrut
\end{tabular}
\vspace{1mm}
\caption{Comparison of all characterization techniques.}
\label{tab:comps}
\end{table*}

Tables \ref{tab:Measurement-comparison-table} and \ref{tab:Measurement-comparison-tableSFWM} summarize the results of applying the explored techniques to our engineered SPDC source and SFWM sources. The implied purities are consistent across all the techniques, following the same trends as a function of filtering and length of fibre, and exhibiting lower values for the phase-sensitive correlation function and two-source HOM measurements, as expected.

Having performed all the techniques ourselves, we have access to all the experimental parameters and can compare the techniques explored, highlight their relative advantages and disadvantages, and provide metrics that can be used to indicate which techniques one may want to pursue based on experimental constraints. This comparison is provided by Table \ref{tab:comps}. In the first two rows we summarize the capabilities of each technique to resolve the JSI and measure the joint phase. For those techniques capable of resolving the JSI we state the spectral resolution of our implementation, which is based on the size of a histogram bin in each data set. As these values are equipment-dependent, in the next row we state the resource that limits the resolution. We list the background-subtracted peak count rates, whether singles or coincidence counts. The nominal acquisition time is the acquisition time per bin multiplied by the number of bins; in practice, the overall acquisition time may be longer due to the finite time required to perform computation and/or mechanical adjustments. Taking the nominal acquisition time and dividing by the spectral resolution gives an approximate acquisition time per bin for our implementations. The `Raw SNR' for counting measurements is the signal-to-noise ratio (SNR) defined in the context of Poisson statistics as the square root of the maximum number of background-subtracted counts in the measurement. For the stimulated-emission-based measurement, it is determined by taking the mean about the peak value and dividing by the standard deviation. However, this raw signal-to-noise ratio  fails to capture the relative ease of data collection (for example, the scanning monochromator measurement might in principle be capable of yielding just as clear a signal as the stimulated measurement, but we would need to use a monochromator with higher resolution and would need to wait a very long time to achieve the same number of counts); thus, we normalize the raw SNR by the product of the acquisition time per bin and the spectral resolution. We present this value as the `Scaled SNR' in the last row of Table \ref{tab:comps}. In the case of the correlation function and HOM measurements, we scale by the total acquisition time. From this comparison we can see that the stimulated-emission-based measurement has the highest scaled SNR. However, this technique requires more resources than the correlation function measurements, which were particularly helpful due to the relative simplicity of the measurement.

The optimal technique for a given application will be determined by the resources available and the requirements of the application. Given these constraints, some techniques are easier to implement than others and some provide more information than others. From our experience, performing the correlation function measurement is a good place to start in characterizing the correlations in a photon-pair source. It provides the purity of the heralded single-photon state with minimal required equipment, and can serve as an initial diagnostic before more detailed investigations in particular degrees of freedom. If one then wishes to gain a detailed, high-resolution, high signal-to-noise measurement of the joint spectrum, and a tunable coherent light source with sufficiently narrow bandwidth is available, the stimulated emission tomography technique is a relatively straightforward measurement that does not require long integration times or significant analysis procedures subsequent to measurement. The dispersive-fibre method and diagonal Fourier-transform method provide good signal-to-noise ratio measurements of the joint spectral intensity. For the dispersive-fibre method a time-to-digital converter is required and the wavelengths of the photons must be such that they undergo sufficient dispersion in available fibre to provide the desired frequency resolution. The diagonal Fourier-transform method is suitable if the joint spectrum is approximately Gaussian and only the joint spectral intensity major and minor axes are required. The monochromator measurement is the most time-consuming technique and has poor signal-to-noise ratio, but may be suitable in the case of a high brightness source, when it is desirable to measure the joint spectral intensity, and only the capabilities of single-photon detection and sufficient resolution spectral filtering are available. The Hong-Ou-Mandel measurement is more difficult than the correlation function measurement as it relies on four-fold coincidence detection. It is usually employed specifically when one wishes to demonstrate the indistinguishability of two sources; as such it is a benchmark for source validation. We hope that our present work will provide a useful description and comparison of techniques available for joint spectral intensity and single heralded-photon purity measurements.

\section{Acknowledgements}

We thank Thomas Gerrits for helpful discussions and a loan of equipment, and Alan Migdall and Offir Cohen for helpful discussions.
Funding for this work has been provided in part by NSF Grant No. PHY 09-03865 ARRA, NSF Grant Nos. 1521110 and 1640968, by Office of Naval Research (ONR) Grant No. N00014-13-1-0627, by US Army ARO DURIP Grant No. W911NF-12-1-0562, by CONACYT, Mexico, by DGAPA (UNAM) and by AFOSR grant FA9550-13-1-0071.


\end{document}